\documentclass[aps,prd,onecolumn,10pt,superscriptaddress,nofootinbib,
preprintnumbers,amsmath,amssymb,floatfix]{revtex4-2}

\usepackage[utf8]{inputenc}
\usepackage[T1]{fontenc}
\usepackage{graphicx}
\usepackage{xcolor}
\usepackage{bm}
\usepackage{hyperref}
\usepackage{orcidlink}
\usepackage{float}

\newcommand{\dd}{\mathrm{d}}

\newcommand{\jump}[1]{\left[#1\right]}
\newcommand{\cN}{\mathcal{N}}
\newcommand{\cD}{\mathcal{D}}
\newcommand{\cQ}{\mathcal{Q}}

\begin{document}

\title{Decoupled asymmetric thin-shell wormholes}

\author{A. López-Hormazábal \orcidlink{0009-0002-8132-3102}}
\email{a.lopez07@ufromail.cl}
\affiliation{Departamento de Ciencias Físicas, Universidad de La Frontera, Casilla 54-D, 4811186 Temuco, Chile.}

\author{Hishan Farfán-Bachiloglu \orcidlink{0000-0002-1382-5605}}
 \email{hishan.farfan@ucn.cl}
\affiliation{Departamento de F\'isica, Universidad Cat\'olica del Norte, Av. Angamos 0610, Antofagasta, Chile.}

\author{Kazuharu Bamba
\orcidlink{0000-0001-9720-8817}}
\email{bamba@sss.fukushima-u.ac.jp}
\affiliation{ Faculty of Symbiotic Systems Science, Fukushima University, Fukushima 960-1296, Japan.}

\author{Francisco Tello-Ortiz \orcidlink{0000-0002-7104-5746}}
\email{francisco.tello@ufrontera.cl}
\affiliation{Departamento de Ciencias Físicas, Universidad de La Frontera, Casilla 54-D, 4811186 Temuco, Chile.}

\begin{abstract}
We construct reflection-asymmetric thin-shell wormholes by joining a Schwarzschild exterior to its minimal geometric deformation, which shares the same mass and redshift function, so that the asymmetry is carried entirely by the gravitational-decoupling sector. The deformed side is a black hole sourced by a scalar field coupled to nonlinear electrodynamics, in its branch with a real scalar. We show that the total radial null energy of shell and bulk obeys an exact identity and becomes positive for a sufficiently strong deformation. The momentum flux across the shell equals the radial null-energy density of the bulk, and retaining it yields the linear stability condition in closed form. We prove that stability with a conventional shell response requires the bulk to violate the radial null energy condition at the throat; the deformation instead enlarges the stable domain and moves the transition radius inside the photon sphere. For a stationary extension built on Kerr, the junction is solved exactly in the spin, which fixes how the angular momentum is shared between shell and bulk and where the shell behaves as a rotating fluid. A free-precession model provides an estimate of the gravitational-wave signal of the throat.
\end{abstract}

\maketitle

\section{Introduction}
\label{sec:introduction}

Traversable wormholes are globally nontrivial geometries connecting two
otherwise separated spacetime regions.  In four-dimensional general
relativity, the flare-out condition forces a violation of the null energy
condition (NEC) in a neighborhood of a smooth throat~\cite{MorrisThorne1988,
VisserBook,LoboBook}.  The cut-and-paste construction of
Visser~\cite{Visser1989,VisserNPB1989} confines all the exotic stress energy
to a timelike hypersurface, whose dynamics is governed by the Darmois--Israel
equations~\cite{Darmois,Israel}.  Thin-shell wormholes are therefore clean
laboratories in which the global geometry and the material properties of the
throat can be separated.  Their linearized radial stability was first analysed
by Poisson and Visser~\cite{PoissonVisser1995} and has since been extended to
charged and cosmological exteriors~\cite{EiroaRomero2004,LoboCrawford2004,
Eiroa2008}, to higher dimensions~\cite{DiasLemos2010}, and to general dynamic
thin shells~\cite{IshakLake2002,LoboCrawford2005,GarciaLoboVisser2012}.
{ Alternative surface equations of state, including Chaplygin and
generalized Chaplygin gases, provide complementary stability analyses
that emphasize the role of the shell constitutive law~\cite{EiroaSimeone2007,Eiroa2009Chaplygin}.
Regular-black hole backgrounds have also been used to examine the response
of thin shells to radial perturbations~\cite{Halilsoy2014Hayward}.}  The
matching of wormhole interiors to exterior vacua through thin
shells~\cite{LemosLoboOliveira2003,Lobo2004,Lobo2005}, and the analogous
treatment of gravastars~\cite{MartinMoruno2012}, have shown how the energy
conditions on the shell constrain the construction.
{ The distinction between negative local surface density and
integrated matter measures has also been explored for non-spherical
thin-shell geometries~\cite{MazharimousaviHalilsoy2014}.}

Most constructions identify two copies of the same exterior.  Reflection
symmetry is not demanded by the junction formalism, and joining different
geometries enlarges the parameter space and changes the mechanical problem of
the throat~\cite{Forghani2018,GarciaLoboVisser2012}.  In the usual
implementation, however, the two sides differ in their masses, charges or
cosmological constants, which mixes the effect of the asymmetry with that of
different asymptotic data.  When the redshift and shape functions of a side
differ, the shell also exchanges momentum with the bulk, a feature emphasized
in the generic analysis of Ref.~\cite{GarciaLoboVisser2012}.
{ In particular, the asymmetric construction and its variable
surface equation of state were already treated systematically in
Ref.~\cite{Forghani2018}; the present work does not claim asymmetry or the
existence of a flux term as new in themselves.  Its distinction is the
controlled comparison of a Schwarzschild exterior with its radial deformed image,
while keeping the redshift function and ADM mass fixed.}

Gravitational decoupling (GD), and in particular its minimal geometric
deformation (MGD) version\footnote{For the extended case see Ref.~\cite{Ovalle:2018gic}}~\cite{Ovalle2008,Casadio2015,Ovalle2017,OvalleBook},
offers a controlled way to produce such an asymmetry.  Originally devised in
the brane-world context~\cite{CasadioOvalleDaRocha2014,CasadioOvalleDaRocha2015EPL,
DaRocha2017,FernandesSilvaDaRocha2018}, the method has since been developed
into a general scheme for anisotropic sources, together with its extended,
inverse and lower-dimensional versions and its relation to the complexity of
self-gravitating systems~\cite{OvalleCasadioDaRochaSotomayor2018,
Contreras2018,ContrerasBargueno2018a,ContrerasBargueno2018b,
ContrerasRinconBargueno2019,CasadioContrerasEtAl2019,ContrerasStuchlik2022}.
It has produced hairy, rotating and field-sourced black
holes~\cite{OvalleEtAl2018BH,OvalleEtAl2018EKG,
OvalleCasadioContrerasSotomayor2021,ContrerasOvalleCasadio2021,
OvalleContrerasStuchlik2021,AriasEtAl2022,AvalosBarguenoContreras2023} and
stellar configurations of fundamental fields and hybrid
stars~\cite{DaRocha2020PRD,CasadioDaRocha2023,daRocha:2021sqd}, and the
geodesics, quasinormal modes and spectroscopy, thermodynamics, information and
entanglement entropies and conformal anomalies of the resulting black holes
have been analysed~\cite{RamosEtAl2021,AvalosContreras2023,
CavalcantiDePaivaDaRocha2022,Guimaraes:2025jsh,dePaiva:2025eux,
Navarro:2026lrf,FernandesSilvaFerreiraMartinsDaRocha2019,DaRochaTomaz2019,
DaRochaTomaz2020,MeertDaRocha2022}.  GD has also been
connected with quantum gravity~\cite{Casadio:2024vfh}, with analog gravity in
fluid flows~\cite{Casadio:2024uwj} and with the holographic description of the
quark--gluon plasma~\cite{daRocha:2024lev}. Also,  MGD has been used to
construct and deform wormhole geometries~\cite{Almatroud2025}.  Here we ask a
sharper question: what thin-shell physics results when the two exteriors glued
at the throat are a black hole and its own MGD image, with the same mass and
the same redshift function?
We answer it for a Schwarzschild seed and an explicit deformed exterior: the
black hole sourced by a scalar field coupled to nonlinear electrodynamics
constructed in Ref.~\cite{GonzalezEspinoza2026}.  Its solution space has two
branches, and we glue the one whose matter sector is best behaved, with a real scalar field and a regular electromagnetic coupling.

The main results are the following.  For a Schwarzschild seed the decoupling
sector reduces to algebraic identities that give the bulk energy conditions
in closed form.  The total radial null-energy quantifier of shell and bulk
obeys an exact identity in which the throat value of the deformation cancels.
The momentum flux across the shell equals the radial null-energy density of the
bulk at the throat, and with it the linear stability condition takes a closed
form that depends on the deformation only through the ratio of this flux to
the surface density.  We prove that stability with a shell response in the
interval $0\le\eta_0<1$ requires the bulk to violate the radial NEC at the
throat, determine how the adopted branch modifies the stability domain, and
show that two profiles with the same slope at the throat have identical linear
stability but different finite-amplitude behaviour.  Finally, for a circular
stationary extension in which the deformed side is built on Kerr, we solve the
junction exactly in the spin, compute the Komar angular momentum and its
distribution between shell and bulk, and determine where the rotating shell
admits an interpretation as a surface fluid.  We then recompute the
Newtonian inertia of the shell and use it in a free-precession model, in the
spirit of Ref.~\cite{Kim2023}, to estimate the gravitational signal of a
precessing throat.
Rotating thin shells assembled from Kerr exteriors were
studied in Ref.~\cite{KasharginSushkov2011}, while counter-rotating
low-dimensional shells were considered in
Ref.~\cite{MazharimousaviHalilsoy2014Rot}.  

The paper is organized as follows.  Section~\ref{sec:mgd} presents the
decoupling sector for a Schwarzschild seed, the deformed exterior and its
energy conditions.  Section~\ref{sec:construction} builds the asymmetric
thin-shell wormhole and establishes the exotic-matter identity.
Section~\ref{sec:stability} contains the radial dynamics, the closed-form
stability condition and its consequences.  Section~\ref{sec:rotation} develops
the stationary extension, its junction and angular-momentum budget, the
inertia of the shell and a free-precession estimate of its gravitational
signal.  Section~\ref{sec:conclusions} summarizes the results, and
Appendix~\ref{app:inertia} contains the expansion of the inertia tensor.
We use signature $(-,+,+,+)$ and units $G=c=1$.  

\section{The deformed exterior}
\label{sec:mgd}

\subsection{Decoupling sector for a Schwarzschild seed}

In MGD the total source is $T^{(m)}_{\mu\nu}+\alpha\theta_{\mu\nu}$ and only the
radial potential of the seed is deformed~\cite{Ovalle2017}.  For the vacuum
seed $A(r)=1-2M/r$,
\begin{equation}
 \dd s^2=-A(r)\dd t^2+\frac{\dd r^2}{B(r)}+r^2\dd\Omega^2,
 \qquad
 B=A+\alpha f,
 \qquad
 f=l A h ,
 \qquad
 B=AH,\quad H\equiv1+\alpha l h ,
 \label{eq:metric}
\end{equation}
where $l$ is the integration constant of the closure and $h(r)$ its radial
profile.  Since $A+rA'=1$ for Schwarzschild, the quasi-Einstein equations of
the decoupling sector collapse to
\begin{equation}
 8\pi\theta^{0}{}_{0}=\frac{l}{r^2}\left[h+(r-2M)h'\right],
 \qquad
 8\pi\theta^{1}{}_{1}=\frac{l h}{r^2},
 \qquad
 8\pi\theta^{2}{}_{2}=\frac{l\,(r-M)\,h'}{2r^2},
 \label{eq:thetasimple}
\end{equation}
which satisfy the separate conservation law
$(\theta^{1}{}_{1})'-\frac{A'}{2A}(\theta^{0}{}_{0}-\theta^{1}{}_{1})
-\frac{2}{r}(\theta^{2}{}_{2}-\theta^{1}{}_{1})=0$ identically.  With
$\alpha\theta^{\mu}{}_{\nu}=\mathrm{diag}(-\theta^{0}_{0},\theta^{1}_{1},\theta^{2}_{2},\theta^{2}_{2})=\mathrm{diag}(-\rho,p_r,p_t,p_t)$, the null and
strong combinations follow without solving anything,
\begin{equation}
 8\pi r^2(\rho+p_r)=-(r-2M)\,H',
 \qquad
 8\pi r^2(\rho+p_t)=-\left[H-1+\tfrac12(r-3M)\,H'\right],
 \qquad
 8\pi r^2(\rho+p_r+2p_t)=M\,H'.
 \label{eq:NECr}
\end{equation}
Outside the horizon the radial NEC is controlled by the sign of $H'$ alone.

The deformed exterior is the one of Ref.~\cite{GonzalezEspinoza2026}, generated
by the linear closure
$\theta^{0}{}_{0}=a\,\theta^{1}{}_{1}+b\,\theta^{2}{}_{2}$, which with
Eqs.~\eqref{eq:thetasimple} integrates to
$h=[(b-2)r+(4-b)M]^{-n}$ with $n=2(a-1)/(b-2)$.  In terms of
\begin{equation}
 r_b\equiv\frac{4-b}{b-2}\,M,
 \qquad
 \varepsilon\equiv\frac{\alpha\ell}{[(b-2)M]^{n}},
 \qquad
 H(r)=1+\varepsilon\left(\frac{M}{r+r_b}\right)^{n},
 \label{eq:Hmodel}
\end{equation}
the deformation is fixed by the falloff exponent $n$, the length $r_b$ and the
amplitude $\varepsilon$, whose sign is that of $\ell\alpha$.  We take
$2<b<4$, so that $r_b>0$, and $a>b/2$, so that $n>1$ and both ends of the
wormhole will have ADM mass $M$.  The source is the Lagrangian
$\mathcal L_\theta=X^{m}-F^{s}V(\phi)$, with
$X=-\tfrac12\nabla_\mu\phi\nabla^\mu\phi$ and $F=-\tfrac14F_{\mu\nu}F^{\mu\nu}$,
for which the closure holds identically when
$m=(a+b-1)/2a$ and $s=(a+b-1)/2(a-1)$.  A canonical scalar,
$m=1$, selects $n=2$ for every $b$; the representative case $b=3$ has $r_b=M$
and
\begin{equation}
 H(r)=1+\frac{\varepsilon M^2}{(r+M)^2}.
 \label{eq:Hcanonical}
\end{equation}

\subsection{Admissibility, energy conditions and the regular branch}

On the deformed side we retain $r\ge a_0>2M$.  The geometry is Lorentzian,
horizon free and regular there if $H>0$ for all $r\ge a_0$; since $H$ in
Eq.~\eqref{eq:Hmodel} is monotonic, this reduces to
$\varepsilon>-[(a_0+r_b)/M]^{n}$, a weaker requirement than the single-horizon
condition of the black hole, $\varepsilon>-(r_b/M)^{n}$.  For $\varepsilon>0$
both are satisfied for every throat.

With $E\equiv H-1$, Eqs.~\eqref{eq:NECr} become
\begin{equation}
\begin{aligned}
 8\pi r^2\rho&=\frac{E\left[(n-1)r-2nM-r_b\right]}{r+r_b},
 &\qquad
 8\pi r^2(\rho+p_r)&=\frac{nE\,(r-2M)}{r+r_b},\\
 8\pi r^2(\rho+p_t)&=\frac{E\left[(n-2)r-3nM-2r_b\right]}{2(r+r_b)},
 &\qquad
 8\pi r^2(\rho+p_r+2p_t)&=-\frac{nME}{r+r_b}.
\end{aligned}
 \label{eq:ECmodel}
\end{equation}
For the scalar--vector source one finds
$\rho+p_r=-2X\partial_X\mathcal L_\theta$ and
$\rho+p_t=2F\partial_F\mathcal L_\theta$.  In the canonical case $n=2$ this
gives
\begin{equation}
 \phi'^{\,2}=-\frac{H'}{8\pi rH}
 =\frac{\varepsilon M^{2}}{4\pi r\,(r+r_b)\left[(r+r_b)^2+\varepsilon M^2\right]},
 \qquad
 2F\,\partial_F\mathcal L_\theta=-\frac{\varepsilon M^2\,(3M+r_b)}{8\pi r^2(r+r_b)^3},
 \label{eq:phiprime}
\end{equation}
so the two branches of Ref.~\cite{GonzalezEspinoza2026} are the two signs of
$\varepsilon$, uniformly in $r$.  For $\varepsilon>0$ the scalar is real,
regular for $r\ge a_0$ and tends to a constant as $\phi_\infty+O(1/r)$; the
price is $\partial_F\mathcal L_\theta<0$, which is the failure of the
odd-parity criterion~\cite{GannoujiBaez2022,BaezGonzalez2023} and, by the same
identity, a tangential NEC violation.  For $\varepsilon<0$ the vector
coupling has the correct sign but $\phi'^2<0$: the scalar is imaginary, the
radial NEC is violated throughout the exterior, and the bulk already supplies
the kind of matter that supports wormholes by itself.  We adopt the branch
$\varepsilon>0$, which we call the regular branch; the name refers to the
reality of the scalar field and to the regularity of the geometry and of the
electromagnetic coupling in the retained region. In it the radial NEC holds
everywhere in the bulk, $H$ decreases monotonically from $H_0>1$ to $1$, and
the energy density is non-negative for $r\ge4M+r_b$; a throat placed beyond
that radius leaves a deformed exterior with $\rho\ge0$ and $\rho+p_r\ge0$.

Steeper profiles improve the effective fluid further.  For $n>2$ and
$\varepsilon>0$, Eqs.~\eqref{eq:ECmodel} give $\rho\ge0$ for
$r\ge(2nM+r_b)/(n-1)$, $\rho-p_r\ge0$ for $r\ge(2nM+2r_b)/(n-2)$,
$\rho+p_r>0$ everywhere, $\rho-p_t>0$ wherever $\rho\ge0$ because $p_t<0$, and
\begin{equation}
 \rho+p_t\ge0
 \quad\Longleftrightarrow\quad
 r\ge r_{\rm NEC}\equiv\frac{3nM+2r_b}{n-2}.
 \label{eq:rNEC}
\end{equation}
Since $r_{\rm NEC}$ exceeds the other two radii, a throat at
$a_0\ge r_{\rm NEC}$ leaves a bulk satisfying the null, weak and dominant
energy conditions, the strong one being violated by the last of
Eqs.~\eqref{eq:ECmodel}.  These profiles correspond to $m<1$, for which a real
action requires $X^m\to-(-X)^m$, and are treated here as effective fluids.

\section{Asymmetric thin-shell wormhole}
\label{sec:construction}

\subsection{Junction conditions and surface sources}

Let ${\cal M}_+$ be Schwarzschild and ${\cal M}_-$ its deformed counterpart,
\begin{equation}
 A_+=A_-=A,\qquad B_+=A,\qquad B_-=AH.
 \label{eq:sides}
\end{equation}
From each manifold we remove $r<a(\tau)$ and identify the boundaries
$\Sigma_\pm=\{r=a(\tau)\}$, whose induced metric is
$\dd s^2_\Sigma=-\dd\tau^2+a^2\dd\Omega^2$~\cite{VisserNPB1989,PoissonVisser1995}.
The first junction condition requires only the intrinsic geometry to agree:
the proper time and the areal radius of the shell are common, while the
coordinate times are fixed on each side separately by
\begin{equation}
 \dot t_\pm=\frac{\sqrt{B_\pm(a)+\dot a^2}}{\sqrt{A(a)B_\pm(a)}} ,
 \label{eq:tdot}
\end{equation}
and need not coincide since $t_\pm$ are unrelated coordinates of different
manifolds.  The unit normals point from the shell into the respective
exteriors,
$n_\mu^{\pm}=\pm\big(-\dot a\sqrt{A/B_\pm},\sqrt{B_\pm+\dot a^2}/B_\pm,0,0\big)$,
and with $\beta_\pm\equiv\sqrt{B_\pm(a)+\dot a^2}$ the extrinsic curvatures are
\begin{equation}
 K^{\theta}{}_{\theta\,\pm}=\pm\frac{\beta_\pm}{a},
 \qquad
 K^{\tau}{}_{\tau\,\pm}=\pm\frac{1}{\beta_\pm}
 \left[\ddot a+\frac{B_\pm A'}{2A}
 +\frac{\dot a^2}{2}\left(\frac{A'}{A}-\frac{B_\pm'}{B_\pm}\right)\right].
 \label{eq:Ktau}
\end{equation}
The Lanczos equations
$8\pi S^i{}_j=-\big(\jump{K^i{}_j}-\delta^i{}_j\jump{K}\big)$, with
$\jump{X}=X_+-X_-$ and $S^i{}_j=\mathrm{diag}(-\sigma,p,p)$, give
\begin{equation}
 \sigma=-\frac{\beta_++\beta_-}{4\pi a},
 \qquad
 p=\frac{1}{8\pi}\sum_{s=\pm}\left\{
 \frac{1}{\beta_s}\left[\ddot a+\frac{B_sA'}{2A}
 +\frac{\dot a^2}{2}\left(\frac{A'}{A}-\frac{B_s'}{B_s}\right)
 \right]+\frac{\beta_s}{a}\right\}.
 \label{eq:sigmadyn}
\end{equation}
The sum of the two square roots is the signature of gluing two exterior
regions.  The construction is traversable for $a>2M$ and $H(a)>0$, and the
areal radius has a minimum at $\Sigma$.

\begin{figure}[t]
 \centering
 \includegraphics[width=0.92\textwidth]{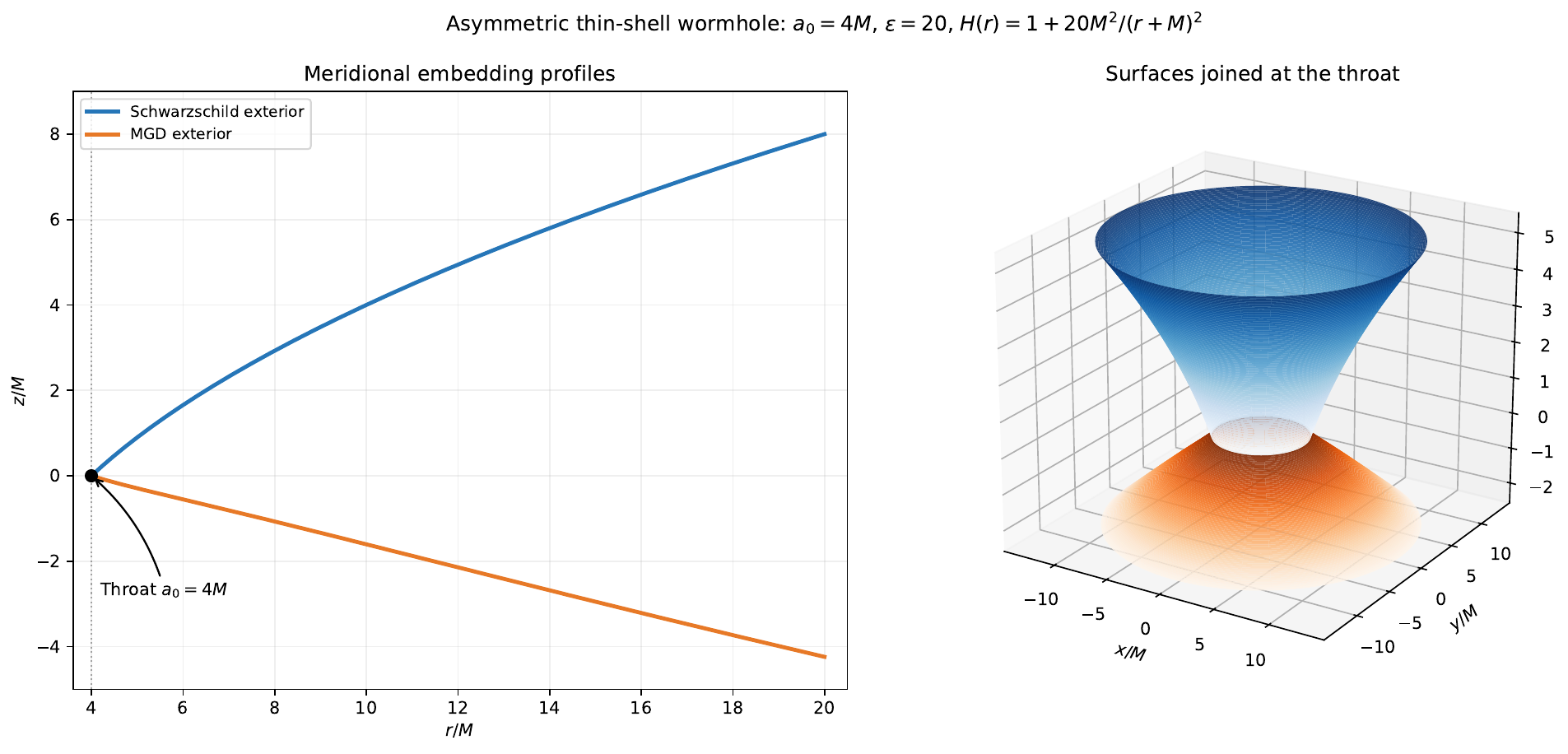}
 \caption{{Euclidean embeddings of constant-time equatorial
 sections of the reflection-asymmetric thin-shell wormhole for $a_0=4M$
 and $\varepsilon=20$.  The Schwarzschild exterior (blue) and the MGD
 exterior (orange) are joined at $r=a_0$.  Their respective radial metric
 functions are $B_+=A$ and $B_-=AH$, where $A=1-2M/r$ and
 $H=1+\varepsilon M^2/(r+M)^2$.  The left panel shows the meridional
 profiles, and the right panel their surfaces of revolution.}}
 \label{fig:embedding}
\end{figure}

{
The reflection asymmetry can also be displayed geometrically by embedding
constant-time equatorial sections of the two static exteriors into Euclidean
three-space.  For $\theta=\pi/2$, their induced two-metrics are
\begin{equation}
 \dd \ell_\pm^2=\frac{\dd r^2}{B_\pm(r)}+r^2\dd\varphi^2,
 \qquad B_+(r)=A(r),\quad B_-(r)=A(r)H(r).
 \label{eq:embeddingmetric}
\end{equation}
A surface of revolution with cylindrical coordinates $(r,\varphi,z_\pm)$
has the Euclidean line element
$\dd\ell^2=[1+(\dd z_\pm/\dd r)^2]\dd r^2+r^2\dd\varphi^2$.
Equating the two line elements gives
\begin{equation}
 z_+(r)=\int_{a_0}^{r}\sqrt{\frac{1}{A(s)}-1}\,\dd s,
 \qquad
 z_-(r)=-\int_{a_0}^{r}\sqrt{\frac{1}{A(s)H(s)}-1}\,\dd s,
 \label{eq:embeddingprofiles}
\end{equation}
with $z_+(a_0)=z_-(a_0)=0$.  The opposite signs distinguish the
outward-directed embeddings of the two exterior regions, joined at the
common throat.  Figure~\ref{fig:embedding} uses the profile
\eqref{eq:Hcanonical} to illustrate how a purely radial MGD deformation
changes the spatial embedding while leaving the redshift function and
asymptotic mass unchanged.  These Euclidean embeddings exist wherever
$0<B_\pm(r)\leq 1$; the condition is satisfied over the plotted range
for the parameters chosen.  The figure represents the static geometry,
not the stationary rotating extension.
}

The matter fields of the decoupling sector live on ${\cal M}_-$ only, and the
shell must source them.  This is implemented by adding to the action the
surface terms $\int_\Sigma\sqrt{-h}\,(A_\mu j_e^{\mu}+\phi\,j_\phi)$, which
produce the jump conditions below.  For the static configuration the vector
equation $\nabla_\mu(\partial_F\mathcal L_\theta F^{\mu\nu})=0$ integrates to
$r^2\sqrt{H}\,\partial_F\mathcal L_\theta\,A_t'=Q$, and with
$2F\partial_F\mathcal L_\theta=\rho+p_t$, $F=\tfrac12HA_t'^{\,2}$,
\begin{equation}
 \partial_F\mathcal L_\theta=\frac{Q^2}{r^4(\rho+p_t)},
 \qquad
 \sigma_e=\Big[n_\mu\,\partial_F\mathcal L_\theta\,F^{\mu\nu}u_\nu\Big]_{\Sigma}=\frac{Q}{a_0^2},
 \label{eq:gauss}
\end{equation}
so the shell carries the whole charge.  For the representative model
$\partial_F\mathcal L_\theta\propto-(r+M)^3/r^2$, which coincides with
$-sF^{s-1}V$ for the electromagnetic invariant and coupling function
reconstructed in Ref.~\cite{GonzalezEspinoza2026}.  The scalar requires a
surface density equal to the jump of its normal derivative,
\begin{equation}
 j_\phi=\Big[n^\mu\partial_\mu\phi\Big]_{\Sigma}=\pm\sqrt{A_0H_0}\,\phi'(a_0),
 \qquad
 j_\phi^{\,2}=(\rho+p_r)\big|_{a_0}=-\frac{A_0H_0'}{8\pi a_0},
 \label{eq:jphi}
\end{equation}
which is real in the regular branch.

\subsection{Static shell and exotic-matter balance}

At $a=a_0$ we write $A_0=1-2M/a_0$, $H_0=H(a_0)$, $q_0=\sqrt{H_0}$ and
$x_0=a_0/M>2$.  Equations~\eqref{eq:sigmadyn} reduce to
\begin{equation}
 \sigma_0=-\frac{\sqrt{A_0}}{4\pi a_0}(1+q_0),
 \qquad
 p_0=-\frac{\sigma_0}{2}\,\frac{a_0-M}{a_0-2M},
 \qquad
 \sigma_0+p_0=\frac{\sigma_0}{2}\,\frac{a_0-3M}{a_0-2M}.
 \label{eq:sigma0}
\end{equation}
The ratio $p_0/\sigma_0=-(x_0-1)/[2(x_0-2)]$ is independent of the
deformation, because the common redshift function makes the deformation
rescale both static curvature components on the minus side by $q_0$.  The
density is negative for every admissible throat, so the weak and dominant
energy conditions fail on the shell, whereas the surface NEC,
$\sigma_0+p_0\ge0$, holds for $2M<a_0\le3M$ and is saturated at the photon
sphere independently of $H$~\cite{Lobo2004,Lobo2005}.

The surface density and the surface NEC are different indicators, and the
volume combinations of the bulk are a third.  To compare the shell and the
bulk on the same footing we use the radial combination, in the spirit of the
volume-integral quantifier of Ref.~\cite{VisserKarDadhich2003}, for which a
thin shell contributes $T^{\mu}{}_{\nu}=S^{\mu}{}_{\nu}\delta(\ell)$ with vanishing
radial pressure and $\ell$ the proper normal distance.  The radial quantifier
is a specific combination defined on the static slice, with its proper volume
element, rather than a universal invariant measure of exotic matter:
\begin{equation}
 \cQ\equiv\int(\rho+p_r)\,\dd V
 =4\pi a_0^2\sigma_0
 +\int_{a_0}^{\infty}\frac{4\pi r^2(\rho+p_r)}{\sqrt{A H}}\,\dd r ,
 \label{eq:Qdef}
\end{equation}
the first term from $\Sigma$ and the second from ${\cal M}_-$, since
${\cal M}_+$ is vacuum.  By Eq.~\eqref{eq:NECr} the bulk integrand is
$-\sqrt{r(r-2M)}\,(\sqrt{H})'$.  Integration by parts produces the throat term
$\sqrt{a_0(a_0-2M)}\,(q_0-1)$, which cancels the explicit $q_0$ dependence of
the shell term, and a term at infinity that vanishes because
$r(\sqrt H-1)\to0$ for $n>1$.  The result is the identity
\begin{equation}
 \cQ=\cQ_{\rm Schw}
 +\int_{a_0}^{\infty}\left[\sqrt{H(r)}-1\right]
 \frac{r-M}{\sqrt{r(r-2M)}}\,\dd r,
 \qquad
 \cQ_{\rm Schw}=-2\sqrt{a_0(a_0-2M)} ,
 \label{eq:Qexact}
\end{equation}
valid for any MGD profile with that falloff.  The explicit boundary
dependence on $H_0$ cancels, which does not make $\cQ$ independent of the
deformation: the integral retains a functional dependence on the whole profile
$H(r)$ on $[a_0,\infty)$; for a monotonic
profile the sign of $\cQ-\cQ_{\rm Schw}$ is the sign of $H-1$.

\begin{figure}[t]
 \centering
 \includegraphics[width=0.52\textwidth]{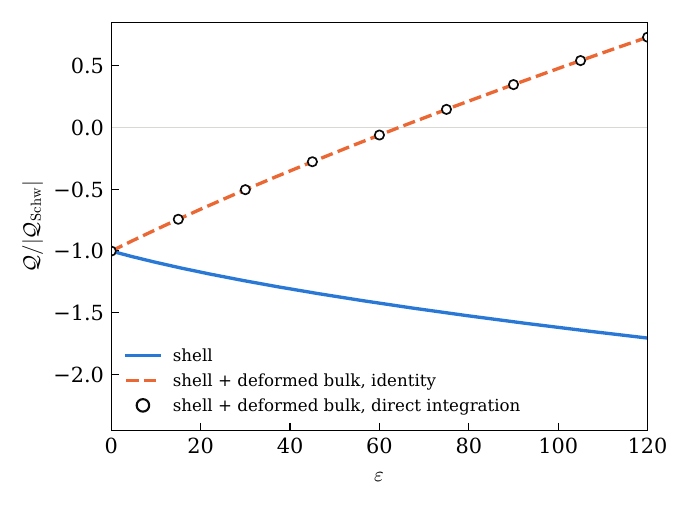}
 \caption{Radial null-energy quantifiers in units of the symmetric value
 $|\cQ_{\rm Schw}|$, as functions of the deformation strength for the
 representative model~\eqref{eq:Hcanonical} with $a_0=4M$.  Solid: shell
 contribution $4\pi a_0^2\sigma_0$.  Dashed: shell plus deformed bulk from the
 identity~\eqref{eq:Qexact}.  Circles: direct evaluation of the
 definition~\eqref{eq:Qdef}.  The total quantifier vanishes at
 $\varepsilon=64.3$.}
 \label{fig:exotic}
\end{figure}

For the regular branch the two contributions move in opposite directions.
The shell becomes more exotic, $|4\pi a_0^2\sigma_0|$ being
$(1+q_0)/2>1$ times the symmetric value, while the bulk, whose
$\rho+p_r$ is positive, more than compensates.  At first order in the
deformation Eq.~\eqref{eq:Qexact} can be integrated in closed form for the
representative model, with $r-M=M\cosh\psi$ and $t_0=\tanh(\psi_0/2)=\sqrt{A_0}$,
\begin{equation}
 \cQ-\cQ_{\rm Schw}=\frac{\varepsilon M}{2}\Big[G(1)-G\big(\sqrt{A_0}\big)\Big]+O(\varepsilon^2),
 \qquad
 G(t)=\frac{4t}{3(3-t^2)}-\frac{1}{3\sqrt3}\ln\frac{\sqrt3+t}{\sqrt3-t},
 \label{eq:Qlinear}
\end{equation}
which gives $0.1290\,\varepsilon M$, $0.1015\,\varepsilon M$ and
$0.0718\,\varepsilon M$ at $a_0=3M$, $4M$ and $6M$.  At finite deformation
$\cQ$ changes sign at $\varepsilon=30.2$, $64.3$ and $161.5$ for the same
throats.  Beyond these values the radial null energy of the configuration as a
whole is positive.  This statement concerns the radial combination only: the
shell density remains negative and the tangential NEC remains violated in the
bulk, so the exotic character of the configuration is redistributed rather
than removed.  Figure~\ref{fig:exotic} shows both quantifiers together with the
direct evaluation of Eq.~\eqref{eq:Qdef}.

\section{Radial dynamics and linear stability}
\label{sec:stability}

The analysis of this section concerns the static, spherically symmetric
wormhole of Sec.~\ref{sec:construction} and perturbations that preserve its
symmetry, with the two exteriors held fixed, in the framework of
Refs.~\cite{PoissonVisser1995,LoboCrawford2004,Eiroa2008,GarciaLoboVisser2012}.

\subsection{Equation of motion and momentum flux}

The first of Eqs.~\eqref{eq:sigmadyn} contains $\dot a$ only through
$\beta_\pm$.  With $S(a)=-4\pi a\sigma(a)>0$, it gives $\beta_++\beta_-=S$, and
since $\beta_+^2-\beta_-^2=B_+-B_-\equiv\Delta$, also
$\beta_+-\beta_-=\Delta/S$.  Adding the squares,
\begin{equation}
 \dot a^2+V(a)=0,
 \qquad
 V(a)=\overline B-\frac{S^2}{4}-\frac{\Delta^2}{4S^2},
 \qquad
 \overline B=\frac{A}{2}\left(1+H\right),
 \quad
 \Delta=A\,(1-H),
 \label{eq:potential}
\end{equation}
where $\overline B=(B_++B_-)/2$.  This is the asymmetric generalization of the
Poisson--Visser potential~\cite{PoissonVisser1995,Forghani2018}: the last
term, quadratic in the asymmetry, vanishes for $H=1$.  Equation~\eqref{eq:potential}
determines the motion once $\sigma(a)$ is known, and $\sigma(a)$ is fixed by the
conservation of the surface stress tensor.

The contracted Gauss--Codazzi equations give
$D_jS^{j}{}_{i}=-\jump{T_{\mu\nu}e^{\mu}_{(i)}n^{\nu}}$, where $e^\mu_{(i)}$ are the
tangent vectors of $\Sigma$~\cite{Israel,GarciaLoboVisser2012}.  Its $\tau$
component, with $e^\mu_{(\tau)}=u^\mu=(\dot t_\pm,\dot a,0,0)$, involves
$T_{\mu\nu}u^\mu n^\nu$, which vanishes on ${\cal M}_+$.  On ${\cal M}_-$, with
$T^{t}{}_{t}=-\rho$ and $T^{r}{}_{r}=p_r$ and the normal of
Sec.~\ref{sec:construction},
\begin{equation}
 T_{\mu\nu}u^\mu n^\nu\big|_{-}
 =-\rho\,u^tn_t+p_r\,u^rn_r
 =\mp\,\dot a\,(\rho+p_r)\,\frac{\beta_-}{B_-} ,
 \label{eq:Tun}
\end{equation}
so only the radial null combination of the bulk crosses the shell.  Dividing
by $\dot a$, the conservation law reads
\begin{equation}
 \sigma'=-\frac{2}{a}(\sigma+p)+\Xi,
 \qquad
 \Xi=(\rho+p_r)\big|_{a}\,\frac{\beta_-}{B_-}
 =-\frac{H'}{8\pi aH}\sqrt{AH+\dot a^2},
 \label{eq:conservationfull}
\end{equation}
where the second form uses Eq.~\eqref{eq:NECr}.  The first term is the
adiabatic work of the surface pressure, and $\Xi$ is the momentum flux
delivered by the bulk.  It is not an additional assumption on the shell
matter: it vanishes only where $\rho+p_r=0$, which for a radial MGD means
$H'=0$, a condition incompatible with a nontrivial asymptotically flat
deformation.  Equivalently, writing the deformed metric as
$-e^{2\Phi}B\,\dd t^2+\dd r^2/B+r^2\dd\Omega^2$ with $\Phi=-\tfrac12\ln H$, the flux
is $\Xi=\Phi'\beta_-/(4\pi a)$, the form given in
Ref.~\cite{GarciaLoboVisser2012}.  In the regular branch $\rho+p_r>0$ and the
flux is positive.

Since $\beta_\pm=(S\pm\Delta/S)/2$, the flux is a function of $a$ and $\sigma$,
and Eq.~\eqref{eq:conservationfull} with a surface equation of state
$p=p(\sigma,a)$ is a closed first-order equation for $\sigma(a)$.  We
characterize the equation of state near equilibrium by
\begin{equation}
 \eta_0=\frac{\partial p}{\partial\sigma}\bigg|_0,
 \qquad
 \gamma_0=-\frac{\partial p}{\partial a}\bigg|_0,
 \label{eq:etagamma}
\end{equation}
with $\gamma_0=0$ for a barotropic shell.  For ordinary matter $\eta_0$ would
be the squared speed of sound, which motivated the interval $0\le\eta_0\le1$
in the early analyses~\cite{PoissonVisser1995}.  For exotic shell matter the
microscopic meaning of $\eta_0$ is not settled, and we use it as a response
coefficient, with $0\le\eta_0<1$ as the conventional benchmark.

At the static radius the stresses~\eqref{eq:sigma0} satisfy $V(a_0)=0$ by
construction, and $V'(a_0)=0$ by virtue of the flux: with
$\mathcal G(a)\equiv\sqrt{B_+}+\sqrt{B_-}=\sqrt A\,(1+q)$ and $q\equiv\sqrt H$,
one finds $V'(a_0)=2\sqrt{B_+B_-}\,(\mathcal G'-S')/\mathcal G$, and
$S_0'=\mathcal G_0'$ holds only when $\Xi_0$ is included.  If the flux is dropped
while $\sigma_0$, $p_0$ and the equation of state are kept,
\begin{equation}
 V'(a_0)\Big|_{\Xi\to0}=\frac{A_0H_0'}{1+q_0}\neq0,
 \label{eq:noflux}
\end{equation}
and the configuration that would be perturbed in that approximation is not an
equilibrium.  A natural measure of the flux is its ratio to the surface
density,
\begin{equation}
 \Lambda_0\equiv\frac{a_0\,\Xi_0}{\sigma_0}
 =\frac{a_0\,q_0'}{1+q_0}
 =\frac{a_0H_0'}{2q_0(1+q_0)} ,
 \label{eq:Lambdadef}
\end{equation}
which is the only combination of the deformation that will appear in the
stability condition.  Since $\sigma_0<0$, $\Lambda_0$ has the sign opposite to
that of $\rho+p_r$ at the throat.

\subsection{Closed-form stability condition}

A small departure $\delta a=a-a_0$ from equilibrium obeys
$\ddot{\delta a}+\tfrac12V''(a_0)\,\delta a=0$, so the throat is linearly stable
when $V''(a_0)>0$ and then oscillates with proper frequency
$\omega_\tau^2=V''(a_0)/2$.  To compute $V''(a_0)$ we use
$\beta_\pm^2=B_\pm-V$, which turns the first of Eqs.~\eqref{eq:sigmadyn} into
$S=\sqrt{B_+-V}+\sqrt{B_--V}$.  Two derivatives evaluated at an equilibrium,
where $V=V'=0$, give
\begin{equation}
 V''(a_0)=\frac{2\sqrt{B_+B_-}}{\mathcal G}\left(\mathcal G''-S''\right)\Big|_{a_0}.
 \label{eq:VppG}
\end{equation}
The second derivative of $S$ follows from Eq.~\eqref{eq:conservationfull} and
the definitions~\eqref{eq:etagamma},
\begin{equation}
 S''=8\pi\eta_0\sigma'-\frac{8\pi}{a}(\sigma+p)-8\pi\gamma_0-4\pi a\,\Xi' .
 \label{eq:Spp}
\end{equation}
On the static profile $4\pi a\Xi=-\sqrt A\,q'$, and because $V'(a_0)=0$ the
derivative of $\beta_-$ along the trajectory coincides with that of
$\sqrt{B_-}$ at $a_0$; hence $(4\pi a\Xi)'=-(\sqrt A\,q')'$ there.  The term
$\sqrt A\,q''$ produced by this derivative cancels exactly the term
$\sqrt A\,q''$ in $\mathcal G''$, and the second derivative of the deformation
drops out. Collecting terms we obtain the following result,
\begin{equation}
 V''(a_0)=\frac{2\sqrt{A_0}\,q_0}{1+q_0}
 \left\{(1+q)\big(\sqrt{A}\big)''+\big(\sqrt{A}\big)'q'
 +(1+2\eta_0)\left[\frac{\sqrt{A}\,(1+q)}{a}\right]'+8\pi\gamma_0\right\}_{a_0},
 \label{eq:Vppgeneral}
\end{equation}
and for the Schwarzschild seed, using $(\sqrt A)'=M/(a^2\sqrt A)$ and
$(\sqrt A)''=-M(2a-3M)/(a^4A^{3/2})$,
\begin{equation}
 V''(a_0)=\frac{2q_0}{M^2x_0^{3}(x_0-2)}
 \Big[\cN_0-2\eta_0\,(x_0-2)\,\cD_0\Big]
 +\frac{16\pi\gamma_0\,q_0\sqrt{A_0}}{1+q_0},
 \label{eq:Vppclosed}
\end{equation}
\begin{equation}
 \cN_0=-(x_0^2-3x_0+3)+(x_0-2)(x_0-1)\Lambda_0,
 \qquad
 \cD_0=(x_0-3)-(x_0-2)\Lambda_0 .
 \label{eq:NDLambda}
\end{equation}

The structure of Eq.~\eqref{eq:Vppgeneral} is transparent.  The first two
terms describe the curvature of the background as felt by the shell; the third
is the response of the shell matter, weighted by $1+2\eta_0$, to the slope of
the static density, since
$\sqrt A(1+q)/a=-4\pi\sigma$ along the family of static configurations; the
last term is the explicit radius dependence of the equation of state, which
always acts in favour of stability for $\gamma_0>0$.  In
Eq.~\eqref{eq:Vppclosed}, $\cD_0$ is proportional to the slope of the
static density, $\sigma_0'=(1+q_0)\cD_0/[4\pi M^2x_0^3\sqrt{A_0}]$, and $\cN_0$
collects the background curvature and the flux.  For $H\to1$ one has
$\Lambda_0\to0$ and Eqs.~\eqref{eq:Vppclosed}--\eqref{eq:NDLambda} reduce to the
result of Poisson and Visser~\cite{PoissonVisser1995}; for $\eta_0=0$ they give
$V''=2q_0\cN_0/[M^2x_0^3(x_0-2)]$.  We have checked
Eqs.~\eqref{eq:Vppclosed}--\eqref{eq:NDLambda} symbolically against the direct
second derivative of Eq.~\eqref{eq:potential}.

Two general features follow.  First, the throat physics depends on the
closure hierarchically: $H_0$ fixes the static stresses, $(H_0,H_0')$ fix the
flux and, through $\Lambda_0$, the linear stability, while the whole profile on
$[a_0,\infty)$ fixes admissibility, the total exotic content and the
finite-amplitude motion.  Second, for a barotropic shell the stability
condition $V''(a_0)>0$ reads
\begin{equation}
 \eta_0<\eta_{\rm c}\quad(\cD_0>0),
 \qquad
 \eta_0>\eta_{\rm c}\quad(\cD_0<0),
 \qquad
 \eta_{\rm c}\equiv\frac{\cN_0}{2(x_0-2)\,\cD_0},
 \label{eq:etac}
\end{equation}
while at $\cD_0=0$ the condition does not involve $\eta_0$ and reduces to
$\cN_0>0$; the division defining $\eta_{\rm c}$ is never used there.  The
radius $a_\ast$ where $\cD_0$ vanishes separates the two regimes.  It is the
photon sphere for $\Lambda_0=0$, where the static density is extremal, and it
moves when the flux is present.  As a function of $\Lambda_0$, $\eta_{\rm c}$ is
strictly decreasing on each side of $\cD_0=0$, since
$\partial\eta_{\rm c}/\partial\Lambda_0=-x_0/(2\cD_0^2)$.

Now we define
\begin{equation}
 \Lambda_{1}(x)\equiv\frac{3x^2-13x+15}{(x-2)(3x-5)},
 \qquad
 \Lambda_{N}(x)\equiv\frac{x^2-3x+3}{(x-2)(x-1)},
 \label{eq:thresholds}
\end{equation}
which satisfy $(x-3)/(x-2)<\Lambda_1<\Lambda_N$ and $\Lambda_1>0$ for all $x>2$.
So, using the previous result with $\gamma_0=0$: (i) a linearly stable
throat with some $\eta_0\in[0,1)$ exists if and only if
$\Lambda_0>\Lambda_1(x_0)$, the window being $\max(0,\eta_{\rm c})<\eta_0<1$;
(ii) it is stable for every $\eta_0>0$ if and only if
$\Lambda_0\ge\Lambda_N(x_0)$; (iii) if $\Lambda_0\le\Lambda_1(x_0)$ every
stable configuration has $\eta_0<0$ or $\eta_0>1$.  Since $\Lambda_1>0$ and
$\sigma_0<0$, a window requires $\Xi_0<0$, that is, $\rho+p_r<0$ at the throat:
the bulk must violate the radial NEC there, and for the scalar--vector source
the scalar is imaginary by Eq.~\eqref{eq:jphi}.

The above implies that the inequalities among $\Lambda_1$, $\Lambda_N$ and
$(x-3)/(x-2)$ follow from
$\Lambda_1-(x-3)/(x-2)=x/[(x-2)(3x-5)]$ and
$\Lambda_N-\Lambda_1=2x/[(x-1)(3x-5)]$, and $\Lambda_1>0$ because $3x^2-13x+15$
has no real roots.  Direct substitution gives $\eta_{\rm c}=1$ at
$\Lambda_0=\Lambda_1$ and $\eta_{\rm c}=0$ at $\Lambda_0=\Lambda_N$.  For
$\Lambda_0>\Lambda_1$ one has $\cD_0<0$, so stability requires
$\eta_0>\eta_{\rm c}$, and monotonicity gives $\eta_{\rm c}<1$; this proves (i)
and, together with $\eta_{\rm c}\le0$ for $\Lambda_0\ge\Lambda_N$, (ii).  For
$\Lambda_0\le\Lambda_1$ there are three cases: if $\cD_0>0$ then
$\Lambda_0<\Lambda_N$ and $\cN_0<0$, so $\eta_{\rm c}<0$; if $\cD_0<0$ then
monotonicity gives $\eta_{\rm c}\ge1$ and stability requires $\eta_0>\eta_{\rm c}$;
if $\cD_0=0$ then $\cN_0<0$ and the throat is unstable.  This proves (iii).
\hfill$\square$

The above statements hold for any radial MGD deformation of Schwarzschild, whatever
the closure, and it gives the mechanical meaning of the windows: the
momentum flux must be directed so as to make the slope of the static density
positive while keeping the curvature term small, and this requires negative
radial null energy to cross the shell.  The symmetric wormhole, with
$\Lambda_0=0$, is the familiar case without such
windows~\cite{PoissonVisser1995,Eiroa2008}, and the obstruction persists for
every deformation whose bulk satisfies the radial NEC at the throat.  In the
model~\eqref{eq:Hmodel} the windows would require $\varepsilon<0$ with
$H_0\lesssim1/4$, far beyond the single-horizon bound of the black hole.

\begin{figure}[t]
 \centering
 \includegraphics[width=0.52\textwidth]{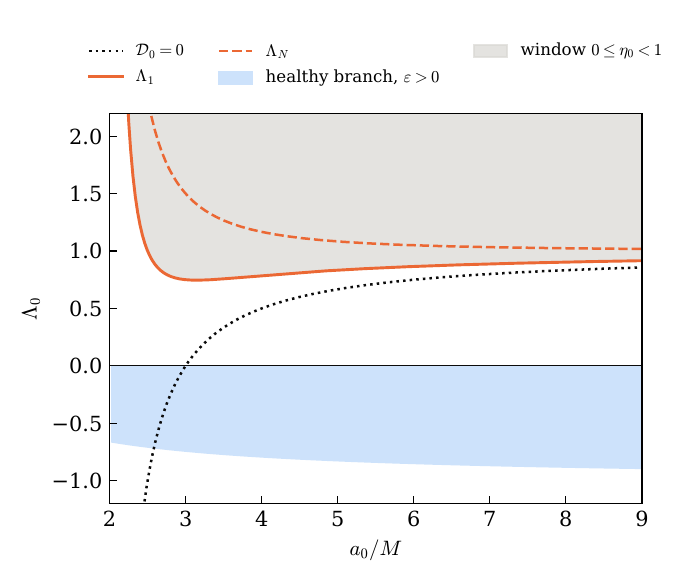}
 \caption{The plane of throat radius and flux parameter $\Lambda_0$ of
 Eq.~\eqref{eq:Lambdadef}.  Dotted: $\cD_0=0$, the transition between the two
 stability regimes.  Solid and dashed: the thresholds $\Lambda_1$ and
 $\Lambda_N$ of Eq.~\eqref{eq:thresholds}.  A window of stability with
 $0\le\eta_0<1$ exists only in the shaded region above $\Lambda_1$.  The band
 below the axis is the range~\eqref{eq:Lambdamodel} spanned by the regular
 branch of the representative model.}
 \label{fig:lambda}
\end{figure}

\begin{figure}[t]
 \centering
 \includegraphics[width=0.92\textwidth]{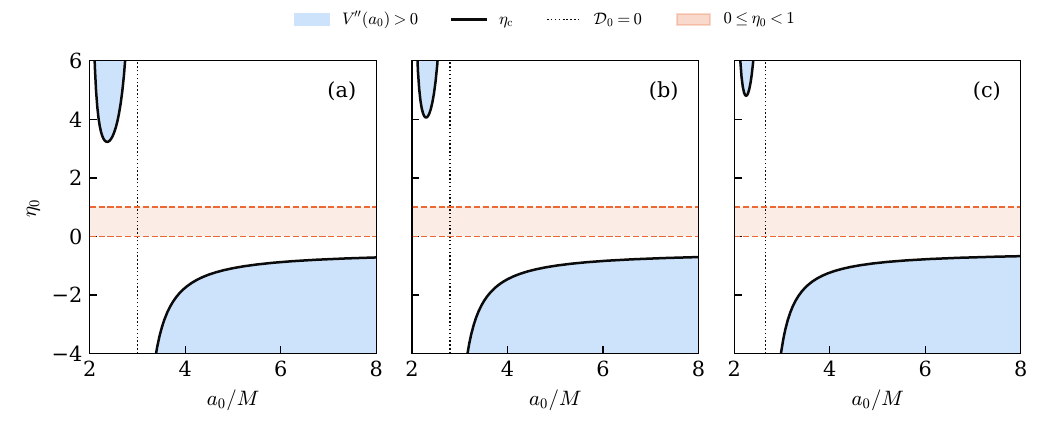}
 \caption{Regions of linear stability, $V''(a_0)>0$, for a barotropic shell in
 the representative model~\eqref{eq:Hcanonical}: (a) symmetric Schwarzschild
 wormhole, $\varepsilon=0$; (b) $\varepsilon=20$; (c) $\varepsilon=200$.  The
 solid curve is $\eta_{\rm c}$, the dotted line the transition radius
 $a_\ast=3M$, $2.79M$ and $2.65M$, and the band marks $0\le\eta_0<1$.}
 \label{fig:stability}
\end{figure}

For the regular branch,
\begin{equation}
 \Lambda_0=\frac{n\,x_0}{2(x_0+x_b)}\left(\frac{1}{q_0}-1\right)\in
 \left(-\frac{n\,x_0}{2(x_0+x_b)},\,0\right),
 \qquad x_b\equiv\frac{r_b}{M},
 \label{eq:Lambdamodel}
\end{equation}
and Fig.~\ref{fig:lambda} shows that this band lies entirely below
$\Lambda_1$.  The deformation nevertheless reshapes the stability domain in
two ways.  The transition radius moves inside the photon sphere: $a_\ast$
decreases monotonically with $\varepsilon$, being $2.91M$, $2.79M$ and
$2.65M$ for $\varepsilon=5$, $20$ and $200$ in the representative model, and
tends to the root of $(x-3)+(x-2)\,n x/[2(x+x_b)]=0$ as
$\varepsilon\to\infty$, which for $n=2$ and $x_b=1$ is $(1+\sqrt{5/2})M$.  And
for $a_0>a_\ast$, where stability requires $\eta_0<\eta_{\rm c}<0$, the
threshold rises with the deformation from its Schwarzschild value towards
the bound
\begin{equation}
 \eta_{\rm c}<-\frac{x_0-1}{2(x_0-2)},
 \label{eq:etabound}
\end{equation}
which is the limit $\Lambda_0\to-\infty$ of Eq.~\eqref{eq:etac}.  At $a_0=4M$,
for example, $\eta_{\rm c}=-1.75$ for Schwarzschild and $-1.46$ for
$\varepsilon=20$.  The stable domain is enlarged, but it never reaches the
interval $0\le\eta_0<1$.  Figure~\ref{fig:stability} displays the regions
$V''(a_0)>0$ evaluated directly from Eq.~\eqref{eq:Vppclosed}, and
Table~\ref{tab:branches} compares the two branches of the deformed exterior
once glued to Schwarzschild.

\begin{table}[htbp]
\caption{Properties of the asymmetric wormhole for the two branches of the
canonical model $n=2$, grouped by type.  Comparisons refer to the symmetric
Schwarzschild wormhole with the same throat radius.}
\label{tab:branches}
\begin{ruledtabular}
\begin{tabular}{llcc}
 & & regular branch, $\varepsilon>0$ & $\varepsilon<0$ \\
\hline
Geometry & throat value $H_0$ & $>1$ & $<1$ \\
 & single-horizon black hole exterior & for all $\varepsilon$ & only for $\varepsilon>-(r_b/M)^2$ \\
Bulk energy conditions & $\rho+p_r$ & $>0$ & $<0$ \\
 & $\rho+p_t$ & $<0$ & $>0$ \\
 & $\rho$ & $\ge0$ for $r\ge4M+r_b$ & $\ge0$ for $r\le4M+r_b$ \\
Field theory & scalar field & real & imaginary \\
Exotic matter & shell content, $|4\pi a_0^2\sigma_0|$ & larger & smaller \\
 & total radial quantifier $\cQ$ & larger, positive for large $\varepsilon$ & more negative \\
Radial stability & momentum flux $\Xi_0$ & $>0$ & $<0$ \\
 & transition radius $a_\ast$ & $<3M$ & $>3M$ \\
 & window $0\le\eta_0<1$ & none & if $\Lambda_0>\Lambda_1(x_0)$ \\
\end{tabular}
\end{ruledtabular}
\end{table}

\subsection{Finite-amplitude motion}
\label{subsec:finite}

Linear stability does not determine the size of the basin.  To probe it we
give the throat a radial velocity $\dot a_0$ at $a=a_0$.  The equation of
motion requires $V(a_0)=-\dot a_0^2$, so the initial density is fixed by the
first of Eqs.~\eqref{eq:sigmadyn},
$\sigma(a_0)=-[\sqrt{A_0+\dot a_0^2}+\sqrt{A_0H_0+\dot a_0^2}]/(4\pi a_0)$, and
the shell obeys the linear barotropic law $p=p_0+\eta_0(\sigma-\sigma_0)$, with
the static values of Eq.~\eqref{eq:sigma0} held fixed for the whole family of
initial data.  Equation~\eqref{eq:conservationfull}, flux included, is
integrated inwards and outwards from $a_0$ and the potential
\eqref{eq:potential} is built from the solution. Small-velocity perturbations of thin-shell wormholes
have also been considered for regular Hayward exteriors~\cite{Halilsoy2014Hayward};
here the finite-amplitude analysis retains the momentum flux associated with
the radial MGD deformation.
The motion is confined to
$V\le0$, and it is an oscillation when two turning points bracket $a_0$ within
the admissible domain.  Along each solution the tangential pressure computed
from the second of Eqs.~\eqref{eq:sigmadyn}, with $\dot a^2=-V$ and
$\ddot a=-V'/2$, reproduces the equation of state to a relative accuracy
better than $10^{-9}$, so the trajectories satisfy the full set of junction
conditions.  The critical velocity $v_{\rm c}$ at which one of the barriers
disappears is stable to six digits under changes of the integration
tolerance and of the sampling of $V$.

\begin{figure}[t]
 \centering
 \includegraphics[width=0.92\textwidth]{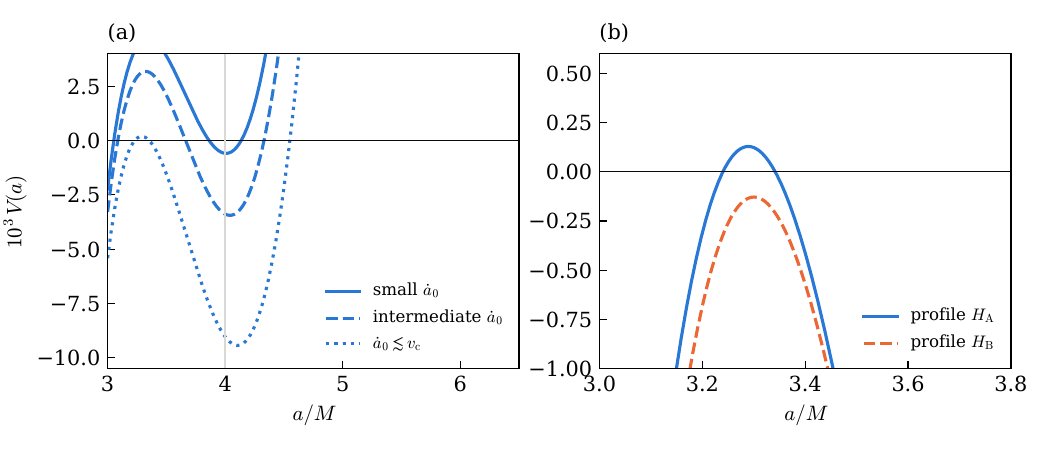}
 \caption{Effective potential of the moving throat for a linear barotropic
 shell with $\eta_0=-2$ and $a_0=4M$; the motion takes place where $V\le0$.
 (a) Representative model with $\varepsilon=20$ and initial velocities
 $\dot a_0=0.024$, $0.058$ and $0.095$, the critical value being
 $v_{\rm c}=0.097$.  (b) Profiles $H_{\rm A}$ and $H_{\rm B}$ with equal
 $H_0$ and $H_0'$ and different $H_0''$ (see text), for $\dot a_0=0.0957$,
 between their critical velocities; the inner barrier exists only for
 $H_{\rm A}$.}
 \label{fig:well}
\end{figure}

Figure~\ref{fig:well}(a) shows the result for $a_0=4M$, $\varepsilon=20$ and
$\eta_0=-2$, a configuration that is linearly stable by
Eq.~\eqref{eq:etac}.  The well deepens with $\dot a_0$ until, at
$v_{\rm c}=0.097$, the inner barrier disappears; just below it the shell
oscillates between $3.30M$ and $4.56M$.  The potential depends on the initial
velocity through $\sigma(a_0)$ and on the chosen equation of state, so the
critical velocity characterizes the pair formed by the exteriors and this
surface-matter prescription.  Figure~\ref{fig:well}(b) tests the
hierarchy of the previous subsection within the same prescription, with
identical $\eta_0$, $p_0$ and $\sigma_0$.  The profile
$H_{\rm B}=1+\varepsilon_{\rm B}[M/(r+r_{\rm B})]^4$ with $r_{\rm B}=6M$ and
$\varepsilon_{\rm B}=8000$ has the same $H_0=1.8$ and $H_0'=-0.32/M$ as the
representative profile $H_{\rm A}$ with $\varepsilon=20$, but
$H_0''=0.160/M^2$ instead of $0.192/M^2$.  The two give the same
$M^2V''(a_0)=0.0637$, but different critical
velocities, $0.0970$ and $0.0944$: for an initial velocity between them the
throat oscillates with $H_{\rm A}$ and collapses with $H_{\rm B}$.

\section{Stationary extension}
\label{sec:rotation}

\subsection{A circular stationary ansatz and its effective source}

Stationary, axisymmetric wormholes are conveniently described by circular
metrics, in which the Killing fields $\partial_t$ and $\partial_\varphi$ are
orthogonal to a family of two-surfaces~\cite{Teo1998}.  Such a metric is not
determined by its static limit: the static exterior of Sec.~\ref{sec:mgd} is
the limit of many stationary families, and generating techniques such as the
Newman--Janis \cite{NewmanJanis1965} and Gürses--Gürsey algorithms~\cite{
GursesGursey1975} or the procedure of Azreg-A\"inou~\cite{AzregAinou2014}
each select one of them, which in general does not solve the field equations
of a given matter sector. We do not use a generating technique.  We prescribe
a circular extension by a physical requirement: both ends must keep the same
mass and the same angular momentum, so that, as in the static case, the
asymmetry is confined to the radial potential.

Denote by $j=J/M$ the Kerr rotation parameter, $\chi=J/M^2$, and
$\Sigma_K=r^2+j^2\cos^2\vartheta$, $\Delta_K=r^2-2Mr+j^2$.  We keep ${\cal M}_+$
as Kerr and take for ${\cal M}_-$ the metric whose components
$g_{tt}$, $g_{t\varphi}$, $g_{\varphi\varphi}$ and $g_{\vartheta\vartheta}$ are those of
Kerr with the same $M$ and $j$, while
\begin{equation}
 g^{rr}_-=\frac{\Delta_K}{\Sigma_K}\,H(r).
 \label{eq:kerransatz}
\end{equation}
This defines a family of circular stationary metrics, exact in $j$, which
is one choice among many with the same static limit.  Its source
$T^{\mu}{}_{\nu}\equiv G^{\mu}{}_{\nu}/8\pi$ is an effective anisotropic tensor; it
does not by itself identify the scalar--vector matter of the static solution,
and whether the scalar--vector action admits rotating fields producing it is a
separate question.  For $H\to1$ the metric is Kerr and the source vanishes;
for $j\to0$ it reduces to Eq.~\eqref{eq:metric} with the source of
Sec.~\ref{sec:mgd}.  Since $a_0>2M\ge r_+$ and the Kerr ergosurface lies at
$r\le2M$, the retained region contains neither horizon nor ergoregion; with
$H>0$ and $H$, $H'$, $H''$ bounded on $[a_0,\infty)$, all metric functions are
smooth there, so the curvature invariants are finite, and the junction
surface $r=a_0$ is timelike.

Since the metric~\eqref{eq:kerransatz} depends only on $r$ and $\vartheta$ and is
invariant under the simultaneous inversion $(t,\varphi)\to(-t,-\varphi)$, the
components $G_{tr}$, $G_{t\vartheta}$, $G_{r\varphi}$ and $G_{\vartheta\varphi}$ vanish
identically. The remaining off-diagonal component of the meridional block
vanishes as well: a direct calculation gives $G_{r\vartheta}=R_{r\vartheta}=0$ for an
arbitrary profile $H(r)$ and to all orders in $j$. The effective source
therefore shares the circular structure of Kerr, and its off-diagonal
components are confined to the $(t,\varphi)$ block. Because $g_{t\varphi}\neq0$,
the covariant and mixed components of this block, $G_{t\varphi}$, $G^{t}{}_{\varphi}$
and $G^{\varphi}{}_{t}$, differ from one another, and we characterize the source
through its components in the orthonormal frame of the zero-angular-momentum
observers (ZAMOs),
\begin{equation}
 e_{(t)}=\frac{\partial_t+\omega\,\partial_\varphi}{N},
 \quad
 e_{(r)}=\sqrt{g^{rr}}\,\partial_r,
 \quad
 e_{(\vartheta)}=\frac{\partial_\vartheta}{\sqrt{\Sigma_K}},
 \quad
 e_{(\varphi)}=\frac{\partial_\varphi}{\sqrt{g_{\varphi\varphi}}},
 \qquad
 \omega=-\frac{g_{t\varphi}}{g_{\varphi\varphi}},
 \quad
 N^2=-g_{tt}+\frac{g_{t\varphi}^2}{g_{\varphi\varphi}},
 \label{eq:zamo}
\end{equation}
in which $T_{(a)(b)}=e_{(a)}^{\mu}e_{(b)}^{\nu}T_{\mu\nu}$ has the diagonal
components $\rho=T_{(t)(t)}$, $p_r$, $p_\vartheta$, $p_\varphi$ and the single
off-diagonal component $T_{(t)(\varphi)}=-T^{(t)(\varphi)}$.  Since the metric is
invariant under $(j,\varphi)\to(-j,-\varphi)$, the diagonal frame components are
even in $j$ and $T^{(t)(\varphi)}$ is odd.  The diagonal components therefore
coincide with those of the static anisotropic fluid of Eqs.~\eqref{eq:NECr} up
to $O(j^2)$, and the leading new quantity is an azimuthal momentum density,
\begin{equation}
 T^{(t)(\varphi)}=\frac{3J\sqrt{A}\,\sin\vartheta}{16\pi r^3}\,H'(r)+O(J^3),
 \qquad
 8\pi T^{t}{}_{\varphi}=\frac{3J\sin^2\vartheta}{2r^2}\,H'(r)+O(J^3).
 \label{eq:pphi}
\end{equation}
The mixed component follows from
$G^{t}{}_{\varphi}=-(\sin^2\vartheta\sqrt{H}/2r^2)(r^4\sqrt{H}\,\omega')'$, valid at
first order with the Kerr frame dragging $\omega=2J/r^3$.  At second order the
diagonal components are corrected and $p_\vartheta\neq p_\varphi$; the Ricci
scalar, for instance, is
\begin{equation}
 \begin{aligned}
 r^2R={}&-2(H-1)-(2r-3M)H'
 +\frac{j^2}{r^3}\Big[2r(H-1)\sin^2\vartheta+4M(H-1)\cos^2\vartheta-(r^2-3r^2\cos^2\vartheta+5Mr\cos^2\vartheta)H'\Big]\\&+O(j^4),
 \end{aligned}
 \label{eq:Ricci}
\end{equation}
which vanishes for $H=1$ and reduces to the static value for $j=0$.

Along $e_{(t)}\pm e_{(r)}$ the combination is $\rho+p_r$, unchanged up to
$O(j^2)$.  Along the azimuthal null directions $k_\pm=e_{(t)}\pm e_{(\varphi)}$,
\begin{equation}
 T_{\mu\nu}k_\pm^{\mu}k_\pm^{\nu}=\rho+p_\varphi\pm2T_{(t)(\varphi)}
 =\rho+p_t\mp2T^{(t)(\varphi)}+O(j^2),
 \label{eq:NECphi}
\end{equation}
so rotation splits the tangential NEC according to the sense of motion, by an
amount of first order in the spin.  In the regular branch $\rho+p_t<0$, and
for slow rotation both azimuthal combinations remain negative: the effective
source of the stationary extension inherits, point by point, the
energy-condition pattern of the static one.

\subsection{Junction, angular momentum and the rotating shell}

The junction surface is $r=a_0$, with intrinsic coordinates
$(t,\vartheta,\varphi)$.  Since the metrics have no $g_{ri}$ components, the unit
normal pointing towards increasing $r$ is $\sqrt{g^{rr}}\,\delta^\mu_r$ on either
side, and for it $K_{ij}=\tfrac12\sqrt{g^{rr}}\,\partial_rg_{ij}$ for all
intrinsic components, including $K_{t\varphi}$.  The induced metrics of the two
sides coincide, so the first junction condition holds, and the extrinsic
curvature of ${\cal M}_-$ computed with this normal is $\sqrt{H_0}$ times that
of Kerr, $K^{\rm Kerr}_{ij}$, because only $g^{rr}$ differs.  In the wormhole
the physical normals point into ${\cal M}_+$ and ${\cal M}_-$ respectively,
that is, towards increasing $r$ on both sheets but with opposite orientation
relative to $\Sigma$; hence $K^+_{ij}=K^{\rm Kerr}_{ij}$,
$K^-_{ij}=-q_0K^{\rm Kerr}_{ij}$ and $\jump{K_{ij}}=(1+q_0)K^{\rm Kerr}_{ij}$.  The
Lanczos equations then give
\begin{equation}
 S_{ij}=-\frac{1}{8\pi}\Big(\jump{K_{ij}}-h_{ij}\jump{K}\Big)
 =\frac{1+q_0}{2}\,S^{\rm KK}_{ij},
 \label{eq:Sijrescaled}
\end{equation}
exactly in $j$ and for every covariant component, where
$S^{\rm KK}_{ij}=-(K^{\rm Kerr}_{ij}-h_{ij}K^{\rm Kerr})/4\pi$ is the surface
stress tensor of the symmetric Kerr--Kerr wormhole~\cite{KasharginSushkov2011};
indices are raised with the common induced metric $h_{ij}$, so the same factor
multiplies the mixed components.  On the shell we use the ZAMO basis of
$\Sigma$, $e_{(t)}=u=N^{-1}(\partial_t+\omega_{\rm Z}\partial_\varphi)$,
$e_{(\vartheta)}$, $e_{(\varphi)}$, with $\omega_{\rm Z}=-h_{t\varphi}/h_{\varphi\varphi}$ and
$N^2=-h_{tt}+h_{t\varphi}^2/h_{\varphi\varphi}$.  The energy density measured by
these observers is
\begin{equation}
\begin{aligned}
 \sigma(\vartheta)&=S_{ij}u^iu^j=-\frac{1+q_0}{2}\,
 \frac{\sqrt{\Delta_0}\left[2a_0^3+j^2a_0+j^2M+j^2(a_0-M)\cos^2\vartheta\right]}
 {4\pi\sqrt{\Sigma_0}\,\Phi_0},\\
 \Phi_0&=a_0^4+j^2a_0^2+2Mj^2a_0+j^2\Delta_0\cos^2\vartheta ,
\end{aligned}
 \label{eq:sigmaKS}
\end{equation}
with $\Delta_0$ and $\Sigma_0$ the Kerr functions at $r=a_0$; for $q_0=1$ this
is the density of Ref.~\cite{KasharginSushkov2011}.  It is a frame component,
not an eigenvalue of $S^{i}{}_{j}$; the two coincide only where the matter is
at rest with respect to the ZAMOs.

\emph{Angular momentum.}  The Komar angular momentum of a closed two-surface
${\cal S}$ is~\cite{Komar1959}
\begin{equation}
 J_{\rm K}[{\cal S}]=\frac{1}{16\pi}\oint_{\cal S}\nabla^\mu\psi^\nu\,\dd S_{\mu\nu},
 \qquad
 \dd S_{\mu\nu}=\left(n_\mu u_\nu-n_\nu u_\mu\right)\dd A ,
 \label{eq:komardef}
\end{equation}
where $\psi=\partial_\varphi$ has period $2\pi$, $u$ is the future unit normal of
the slice $t={\rm const}$, $n$ is the unit normal of ${\cal S}$ within the slice
pointing towards the asymptotic region of the sheet considered, and
$\dd A=\sqrt{g_{\vartheta\vartheta}g_{\varphi\varphi}}\,\dd\vartheta\,\dd\varphi$; with this
orientation $J_{\rm K}=J$ for every sphere of Kerr.  Since $\psi$ is a Killing
field, $\nabla_r\psi_\beta=\tfrac12\partial_rg_{\beta\varphi}$, and for a sphere
$r={\rm const}$ the integrand reduces to
\begin{equation}
 \nabla^\mu\psi^\nu\,\dd S_{\mu\nu}
 =-N\sqrt{g^{rr}}\;g^{t\beta}\,\partial_rg_{\beta\varphi}\;\dd A .
 \label{eq:komarint}
\end{equation}
In Eq.~\eqref{eq:kerransatz} every factor except $\sqrt{g^{rr}}$ is that of
Kerr, so the integrand is $\sqrt{H(r)}$ times the Kerr integrand at every
point of the sphere, and
\begin{equation}
 J_{\rm K}(r)=\sqrt{H(r)}\;J,
 \label{eq:Komar}
\end{equation}
exactly in $j$.  On ${\cal M}_+$ the Komar angular momentum is $J$ for every
sphere.  Consider a slice consisting of the two sheets joined at $\Sigma$,
bounded by the spheres at infinity of ${\cal M}_+$ and ${\cal M}_-$, where
$J_{\rm K}=J$ on both since $H\to1$.  Stokes' theorem applied to the Komar
two-form states that the sum of the two boundary values, $2J$, equals the
angular momentum of everything in between, the shell plus the deformed bulk.
The shell contribution is the sum of the values of $J_{\rm K}$ on the two faces
of $\Sigma$, each evaluated with the normal pointing into its own sheet, and the
bulk contribution is $J-J_{\rm K}(a_0)$:
\begin{equation}
 J_{\rm shell}=(1+q_0)\,J,
 \qquad
 J_{\rm bulk}=(1-q_0)\,J .
 \label{eq:Jbudget}
\end{equation}
The shell value is confirmed independently by the Israel surface current,
\begin{equation}
 J_{\rm shell}=-\oint_\Sigma S_{ij}\,u^i\psi^j\,\dd A=(1+q_0)\,J ,
 \label{eq:Jisrael}
\end{equation}
which we have evaluated with the exact surface stress tensor at finite spin;
both routes agree to all orders in $j$.  The bulk value is the Komar integral
of the effective source.  In the regular branch it is negative; this states
how the total angular momentum is shared, and does not by itself define a
local sense of rotation of the bulk.  The local quantity is the ZAMO momentum
density~\eqref{eq:pphi}, negative for $H'<0$.

\emph{Rotating surface fluid.}  The shell admits an interpretation as a
rotating surface fluid if $S^{i}{}_{j}$ has a timelike eigenvector.  In the ZAMO
basis $S_{(t)(\vartheta)}=S_{(\vartheta)(\varphi)}=0$, and the $(t,\varphi)$ block has the
components $\sigma=S_{(t)(t)}$, $p_\varphi=S_{(\varphi)(\varphi)}$ and
$f=S_{(t)(\varphi)}$.  The eigenvalue problem
$S^{(a)}{}_{(b)}v^{(b)}=\lambda v^{(a)}$, with indices raised by
$\eta={\rm diag}(-1,1)$, gives
$\lambda^2+(\sigma-p_\varphi)\lambda+f^2-\sigma p_\varphi=0$, whose discriminant is
$(\sigma+p_\varphi)^2-4f^2$.  The block has two real eigenvalues with a timelike
eigenvector, and $S^{i}{}_{j}$ is of Hawking--Ellis type~I~\cite{HawkingEllis},
if and only if
\begin{equation}
 (\sigma+p_\varphi)^2>4f^2 ;
 \label{eq:typeI}
\end{equation}
for $(\sigma+p_\varphi)^2<4f^2$ the eigenvalues are complex and the tensor is of
type~IV.  By Eq.~\eqref{eq:Sijrescaled} all three quantities carry the factor
$(1+q_0)/2$, so the domain where Eq.~\eqref{eq:typeI} holds is that of the
Kerr--Kerr shell and is independent of the deformation.  The condition depends
on $\vartheta$ and is imposed at every point of $\Sigma$.  It fails only in a band
of radii around the photon sphere, where $\sigma+p_\varphi$ vanishes while $f$
does not; the band is widest at the equator and narrows towards the poles,
where $f\to0$, so the set of throats for which the fluid interpretation fails
somewhere on $\Sigma$ is the equatorial band.  It is
$2.942M<a_0<3.057M$, $2.882M<a_0<3.113M$ and $2.759M<a_0<3.223M$ for
$\chi=0.05$, $0.1$ and $0.2$, with half-width $6\chi M^2\sqrt{A_0}/a_0$ to leading
order; at $\vartheta=0.3$, for comparison, it reduces to $2.962M<a_0<3.030M$ for
$\chi=0.1$.  Inside the band the geometry is regular and only the fluid
interpretation of the shell is lost.

Outside the band the timelike eigenvector $v=\partial_t+\Omega_{\rm m}\partial_\varphi$
defines the angular velocity of the shell matter.  At first order in the spin
it is independent of $\vartheta$,
\begin{equation}
 \Omega_{\rm m}-\omega_0=-\frac{3JA_0}{a_0^2\,(a_0-3M)},
 \qquad
 \omega_0=\frac{2J}{a_0^3},
 \label{eq:Omegam}
\end{equation}
and of the deformation; the $\vartheta$ dependence, a differential rotation of
the shell, appears at order $j^3$.  The denominator reflects the band: the
expansion is not uniform near $a_0=3M$, and Eq.~\eqref{eq:Omegam} applies for
$|a_0-3M|$ large compared to the band width.  At the same order the
budget~\eqref{eq:Jbudget} and the angular velocity are related by
$J_{\rm shell}=I_{\rm eff}(\Omega_{\rm m}-\omega_0)$, with the relativistic moment of
inertia
\begin{equation}
 I_{\rm eff}=\frac{8\pi}{3}\,\frac{a_0^4\,(\sigma_0+p_0)}{\sqrt{A_0}}
 =-\frac{(1+q_0)\,a_0^3\,(a_0-3M)}{3\,(a_0-2M)} ,
 \label{eq:Ieff}
\end{equation}
which is weighted by the surface enthalpy $\sigma_0+p_0$ of the static shell.
This is an identity at first order in $j$.  For $a_0>3M$, where
$\sigma_0+p_0<0$ and $I_{\rm eff}<0$, the shell matter rotates against the local
inertial frames while carrying positive angular momentum.

\emph{A second extension.}  Demanding instead $T^{t}{}_{\varphi}=0$ on
${\cal M}_-$ gives, at first order, $(r^4\sqrt H\omega')'=0$, so
$\omega_-=6J_-\int_r^{\infty}\dd r'/(r'^4\sqrt{H})$ and the Komar angular momentum
is constant on ${\cal M}_-$.  Continuity of the induced metric then forces
$J_+/J_-=3a_0^3\int_{a_0}^{\infty}\dd r/(r^4\sqrt{H})$, which is smaller than one
in the regular branch, and the shell carries $J_++J_-$.  This shows that the
static limit does not select a unique rotation.  The static results of
Secs.~\ref{sec:construction} and~\ref{sec:stability} are common to both
extensions, as are the first-order matter angular velocity and the type-I
analysis up to the value of $J$ on each side; the
rescaling~\eqref{eq:Sijrescaled}, the equality of the asymptotic angular
momenta and the budget~\eqref{eq:Jbudget} belong to the
ansatz~\eqref{eq:kerransatz}.

\subsection{Inertia of the shell and the radial mode}

Two different notions of inertia can be attached to the rotating shell.  The
relativistic moment of inertia~\eqref{eq:Ieff} is the response of the angular
momentum of the shell to its rotation relative to the local inertial frames.
A Newtonian inertia tensor, of the kind used in Ref.~\cite{Kim2023} to describe
the precession of the Kerr--Kerr shell, is an auxiliary descriptor of the
surface distribution,
$I_{ij}=\int\sigma\,(r^2\delta_{ij}-x_ix_j)\,\dd A$, which requires positions $x_i$
for the points of $\Sigma$ and a choice of area element.  Following
Ref.~\cite{Kim2023} we represent $r=a_0$ as the oblate spheroid
$x+iy=\sqrt{a_0^2+j^2}\,\sin\vartheta\,e^{i\varphi}$, $z=a_0\cos\vartheta$ of the
flat Boyer--Lindquist background, weight it with the ZAMO density
\eqref{eq:sigmaKS}, which is negative, and use the proper area element
$\dd A=\sqrt{h_{\vartheta\vartheta}h_{\varphi\varphi}}\,\dd\vartheta\,\dd\varphi$.  Expanding to
second order in $j$ (Appendix~\ref{app:inertia}),
\begin{equation}
\begin{aligned}
 I_1&=-\frac{2}{3}(1+q_0)\,a_0^3\sqrt{A_0}\left[1+O\!\left(\frac{j^2}{a_0^2}\right)\right],
 \qquad
 I_1-I_3=\frac{1+q_0}{15}\,(6x_0-1)\sqrt{A_0}\;Mj^2,\qquad
 \epsilon_I&\equiv\frac{I_3-I_1}{I_1}=\frac{6x_0-1}{10\,x_0}\,\frac{j^2}{a_0^2}.
\end{aligned}
 \label{eq:inertia}
\end{equation}
The deformation enters $I_1$ and $I_1-I_3$ only through the overall factor
$1+q_0$, which cancels in $\epsilon_I$.  The moments are negative because
$\sigma<0$, and the positive ellipticity reflects the oblate shape of a
distribution of negative energy.  These values depend on the representation:
with the area element of the flat spheroid, used in Ref.~\cite{Kim2023},
$\epsilon_I=3(2x_0-1)j^2/(10x_0a_0^2)$, and other embeddings give other values.
The ellipticity $(13x_0-9)j^2/(20x_0a_0^2)$ quoted in Ref.~\cite{Kim2023}
corresponds to the flat element; the reconstruction of the two integrals in
Appendix~\ref{app:inertia} shows that it differs from the flat-element value
because of the weight of the $\cos^2\vartheta$ part of the density in $I_3$.
The gravitational quadrupole of each asymptotic end is the Geroch--Hansen
moment of Kerr, $-Mj^2$~\cite{Geroch1970,Hansen1974}, independent of the throat;
the shell quadrupole defined by $I_{ij}$ is a different object, and it is not
the radiative quadrupole of either end.

\begin{figure}[t]
 \centering
 \includegraphics[width=0.92\textwidth]{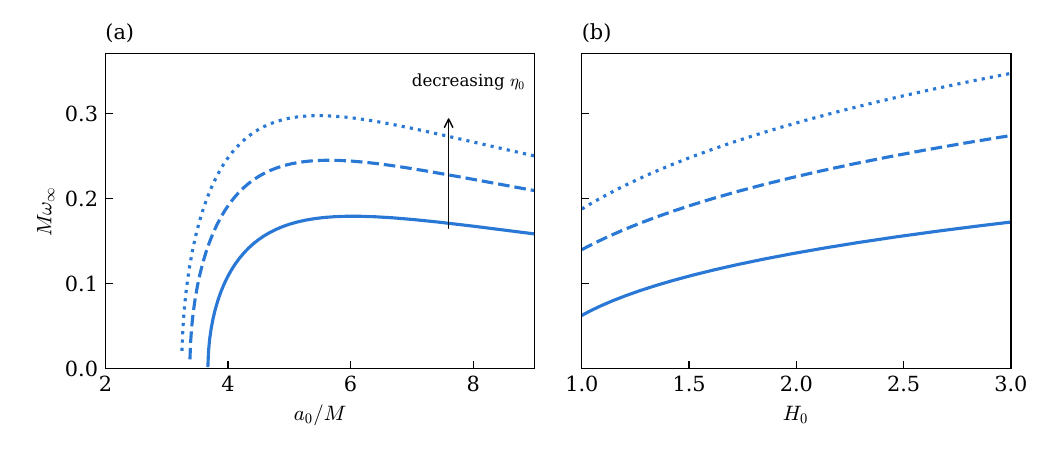}
 \caption{Frequency at infinity of the linear radial mode of the static throat,
 Eq.~\eqref{eq:breathing}, in the regular branch of the representative
 model~\eqref{eq:Hcanonical}, for $\eta_0=-2$ (solid), $-3$ (dashed) and $-4$
 (dotted).  (a) As a function of the throat radius at fixed $H_0=1.5$.
 (b) As a function of $H_0$ at $a_0=4M$.  Curves are drawn where the throat
 is linearly stable.  In physical units
 $f\simeq32.3\,{\rm kHz}\,(M_\odot/M)\,M\omega_\infty$.}
 \label{fig:breathing}
\end{figure}

The radial degree of freedom of the throat is described dynamically by
Sec.~\ref{sec:stability}.  In the static configuration
$\dd t_\pm/\dd\tau=A_0^{-1/2}$ on both sides by Eq.~\eqref{eq:tdot}, since the
redshift function is common, so the proper frequency
$\omega_\tau^2=V''(a_0)/2$ of the linear radial mode is seen at either infinity
as
\begin{equation}
 (M\omega_\infty)^2=\frac{A_0M^2}{2}\,V''(a_0)
 =\frac{q_0}{x_0^{4}}\Big[\cN_0-2\eta_0\,(x_0-2)\,\cD_0\Big],
 \qquad
 f=\frac{\omega_\infty}{2\pi}\simeq32.3\,{\rm kHz}\;
 \frac{M_\odot}{M}\,(M\omega_\infty),
 \label{eq:breathing}
\end{equation}
for a barotropic shell, where $c^3/(2\pi GM_\odot)=32.3$\,kHz restores the
units.  This is the frequency of spherical radial oscillations of the static
throat between fixed exteriors, defined where the throat is stable.  It is
not a polar mode of the rotating system, and a purely spherical pulsation does
not radiate quadrupolar waves; a radiative signal at this frequency would
require its coupling to nonspherical multipoles in the rotating geometry.  In
the regular branch, Fig.~\ref{fig:breathing} gives
$M\omega_\infty\simeq0.06$--$0.35$ for $\eta_0$ between $-2$ and $-4$; for
$a_0=4M$, $\varepsilon=20$ and $\eta_0=-2$, $M\omega_\infty=0.126$, that is
$f=4.1\,{\rm kHz}\,(M_\odot/M)$.

\subsection{A free-precession estimate}
\label{subsec:precession}

The stationary configuration has a timelike Killing field and emits no
gravitational radiation; rotation about the symmetry axis is not radiative.
A precessing throat is a different, nonstationary configuration.  To estimate
its signal we adopt, in the spirit of Ref.~\cite{Kim2023}, the following
phenomenological model, which is an additional dynamical hypothesis rather
than a consequence of the stationary solution: the shell moves as a rigid
axisymmetric body with the inertia tensor~\eqref{eq:inertia} and angular
momentum $L=J_{\rm shell}$; its symmetry axis is tilted by an angle $\alpha$
with respect to $\mathbf L$, no torque acts on it, and the radiation is
computed with the quadrupole formula, neglecting radiation reaction.  The
exteriors are not perturbed self-consistently, so the estimate is not derived
from the linearized field equations and junction conditions.

\begin{figure}[H]
 \centering
 \includegraphics[width=0.92\textwidth]{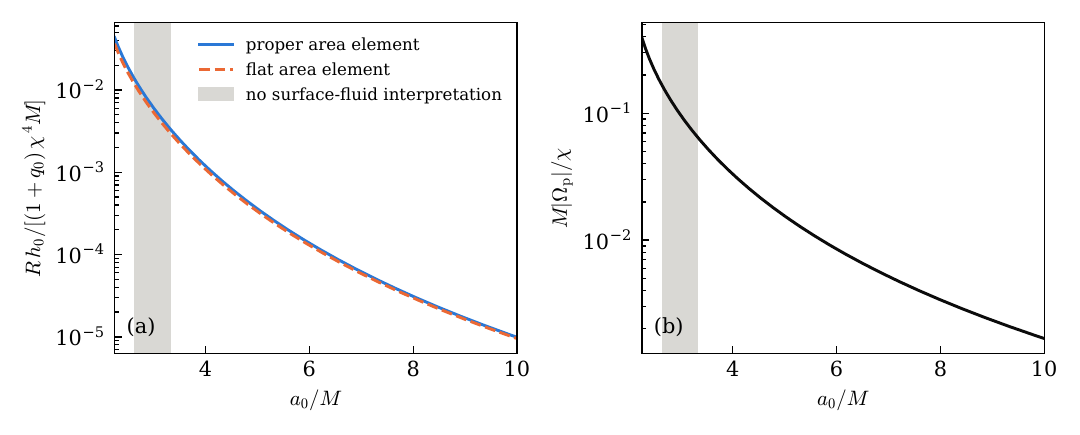}
 \caption{Free-precession estimate of Sec.~\ref{subsec:precession}.
 (a) Amplitude scale $|h_0|$ of Eq.~\eqref{eq:h0}, normalized to
 $(1+q_0)\chi^4M/R$, with the proper (solid) and flat (dashed) area elements.
 (b) Precession rate~\eqref{eq:Omegap} normalized to $\chi/M$.  The shaded band
 is the interval where the surface stress tensor is of type~IV somewhere on
 the shell, shown for $\chi=0.3$.}
 \label{fig:precession}
\end{figure}

For an axisymmetric body the torque-free Euler equations give a symmetry axis
\begin{equation}
 \mathbf n(t)=(\sin\alpha\cos\Omega_{\rm p}t,\ \sin\alpha\sin\Omega_{\rm p}t,\ \cos\alpha),
\end{equation}
which precesses about $\mathbf L$ at the inertial rate
$\Omega_{\rm p}=L/I_1$~\cite{LandauLifshitz}.  With
Eqs.~\eqref{eq:Jbudget} and~\eqref{eq:inertia},
\begin{equation}
 \Omega_{\rm p}=\frac{L}{I_1}=-\frac{3J}{2a_0^3\sqrt{A_0}}\left[1+O\!\left(\frac{j^2}{a_0^2}\right)\right],
 \label{eq:Omegap}
\end{equation}
independent of the deformation; the negative sign, a consequence of $I_1<0$,
means that the axis precesses against $\mathbf L$.  This rate is a property of
the rigid-body model; it is neither the frame-dragging rate $\omega_0$ nor the
matter angular velocity $\Omega_{\rm m}$, and the body-frame rate
$\epsilon_I\,\omega_3$, with $\omega_3$ the spin about the symmetry axis, is
not the frequency seen by a distant observer.  The trace-free quadrupole of
the body,
\begin{equation}
 Q_{ij}=-(I_3-I_1)\left(n_in_j-\tfrac13\delta_{ij}\right),
 \label{eq:Qij}
\end{equation}
rotates rigidly with $\mathbf n(t)$, and the quadrupole formula
$h_{ij}^{\rm TT}=2\ddot Q_{ij}^{\rm TT}/R$ at distance $R$ gives, for an
observer at inclination $\iota$ with respect to $\mathbf L$ and the usual
polarization basis,
\begin{equation}
\begin{aligned}
 h_+&=h_0\Big[2\sin^2\alpha\,(1+\cos^2\iota)\cos2\Omega_{\rm p}t
 -\sin2\alpha\,\sin\iota\cos\iota\,\cos\Omega_{\rm p}t\Big],\\
 h_\times&=2h_0\Big[2\sin^2\alpha\,\cos\iota\,\sin2\Omega_{\rm p}t
 -\tfrac12\sin2\alpha\,\sin\iota\,\sin\Omega_{\rm p}t\Big],
 \qquad
 h_0=\frac{(I_3-I_1)\,\Omega_{\rm p}^2}{R}.
\end{aligned}
 \label{eq:hpc}
\end{equation}
The signal has the harmonics $\Omega_{\rm p}$ and $2\Omega_{\rm p}$, with
amplitudes proportional to $\sin2\alpha$ and $\sin^2\alpha$ respectively, as for
freely precessing rigid bodies~\cite{ZimmermannSzedenits1979,JonesAndersson2002}.
It vanishes for $\alpha=0$, when the body merely spins about its symmetry axis,
and for $j\to0$; for $H\to1$ it reduces to the Kerr--Kerr model.  Restoring
units,
\begin{equation}
 |h_0|=\frac{3\,(1+q_0)\,(6x_0-1)}{20\,x_0^6\sqrt{A_0}}\;\chi^4\,\frac{GM}{c^2R},
 \qquad
 f_{\rm p}=\frac{|\Omega_{\rm p}|}{2\pi}\simeq32.3\,{\rm kHz}\;\frac{M_\odot}{M}\,
 \frac{3\chi}{2x_0^3\sqrt{A_0}} .
 \label{eq:h0}
\end{equation}
The deformation enters only through the factor $(1+q_0)/2$ relative to the
Kerr--Kerr shell, through $J_{\rm shell}$ and $I_{ij}$, while the frequencies do
not depend on it.  Both quantities are shown in Fig.~\ref{fig:precession}.  For
$M=10M_\odot$, $\chi=0.3$, $a_0=4M$ and $\varepsilon=20$, Eq.~\eqref{eq:h0} gives
$f_{\rm p}=32$\,Hz and, at $R=10$\,kpc, $|h_0|=1.1\times10^{-21}$; with
$\alpha=0.1$ the two harmonics reach $1.1\times10^{-22}$ at $\iota=\pi/4$ and
$4.3\times10^{-23}$ at $\iota=0$, respectively.  With the flat area element
the amplitude is smaller by the factor $(6x_0-3)/(6x_0-1)$, about $10\%$ at
$a_0=4M$, which measures the representation dependence of the estimate.

These figures are conditional on the model and set its domain.  The
expansion requires $j/a_0\ll1$ ($j/a_0=0.075$ in the example) and a throat
outside the band where the shell is not a fluid.  The quadrupole formula
assumes a weak field in the generation region, which a throat at
$a_0\sim M$ does not provide, and it ignores backscattering, the coupling to
the effective bulk matter and the presence of two asymptotic regions, each of
which receives radiation.  Radiation reaction is neglected: at fixed $a_0/M$
the fraction of angular momentum radiated per precession cycle scales as
$\chi^7$.  Within these limits Eq.~\eqref{eq:h0} is an order-of-magnitude
estimate of the signal of a precessing throat, not a prediction derived from
the exact solution.

\newpage
\section{Conclusions}
\label{sec:conclusions}

We have constructed thin-shell wormholes whose two sides are a Schwarzschild
exterior and its MGD, with the same mass and the
same redshift function. The asymmetry of the throat is then a property of the
decoupling sector alone, and the deformed side was chosen as the best-behaved
black hole available in the scalar--vector family, with a real scalar field
and a regular electromagnetic coupling.

The construction yields three structural results that do not depend on the
particular closure. The total radial null energy of shell and bulk obeys an
exact identity in which the value of the deformation at the throat drops out,
so that a shell made more or less exotic by the deformation is compensated in
the bulk; for the regular branch the configuration as a whole acquires
positive radial null energy once the deformation is large enough, while its
tangential and surface exotic character remains. The momentum that the bulk
transfers to a moving shell is precisely its radial null-energy density, and
the radial stability of the throat is governed by the ratio of this flux to
the surface density, which makes linear stability blind to the curvature of
the profile while the finite-amplitude basin is not. The stability with a
shell response in the conventional interval demands that negative radial null
energy flow into the shell, so that it cannot be obtained by gluing a
black hole exterior whose bulk satisfies the radial null energy condition;
what such a deformation does instead is enlarge the stable domain and move the
transition between the two stability regimes inside the photon sphere.

For the stationary extension that keeps the same mass and angular momentum on
both sides, the junction is solved exactly in the spin, with the shell being
the Kerr--Kerr shell rescaled by the same factor that governs its static
exotic content. The Komar angular momentum on the deformed side is the Kerr
value weighted by the square root of the deformation, which fixes exactly how
the angular momentum is shared between shell and bulk, and the shell is a
rotating surface fluid everywhere except in a band around the photon sphere
whose width depends on the spin but not on the deformation. The two routes to the
angular momentum of the shell, through the Komar integral and through the
surface current, agree to all orders in the spin.

The Newtonian inertia tensor used in previous estimates of precessional
emission has been recomputed and shown to depend on the representation of the
throat, which corrects the ellipticity quoted before and identifies it as an
auxiliary descriptor. Used within a rigid-body model of free precession, it
gives a signal at the precession rate and its second harmonic whose
frequencies are blind to the deformation, while the amplitude carries the same
rescaling as the surface stresses. The deformation thus leaves on the rotating
throat a clean multiplicative imprint, set by its value at the junction.

Because both exteriors share the mass, the redshift function and, in the
stationary extension, the angular momentum of Schwarzschild or Kerr, the
asymmetric wormhole is locally indistinguishable from a black hole for
observables tied to the exterior geometry alone, such as the photon ring or
orbital dynamics far from the throat. Its imprint must therefore be sought in
the dynamics of the throat itself. A stable throat oscillating radially, or a
precessing rotating shell, would emit at characteristic frequencies fixed by
the mass, the spin and the throat radius. For stellar-mass objects these lie
in the band of the ground-based detectors, from tens of hertz for
precession to kilohertz for radial oscillations. Continuous-wave searches by
LIGO, Virgo and KAGRA, and the higher sensitivity expected from the Einstein
Telescope and Cosmic Explorer, could constrain such signals, while echoes
in the ringdown of compact-object mergers offer a complementary window on
any reflecting structure near the photon sphere. In this model the
deformation leaves the frequencies unchanged and acts on the amplitude
through a single factor set at the junction. Measuring a frequency together
with an amplitude would thus, in principle, separate the geometric
parameters of the throat from the strength of the decoupling sector.
Although these expectations are speculative, they show that thin-shell
wormholes built by GD make definite, falsifiable
statements about the throat that do not rely on any deviation of the
exterior from general relativity.

More broadly, this work establishes GD as a controlled
tool for building asymmetric junctions: it fixes the asymptotic data and the
causal structure of both sides and isolates the physics of the throat in a
single, separately conserved sector. This opens a research line along
several natural directions. The same construction can be applied to charged,
cosmological and regular black hole seeds, and to other closures of the
decoupling sector, to map how the exotic content, the momentum flux and the
stability windows depend on the matter that supports the deformation. Thin
shells joining a deformed exterior to other interiors, such as gravastars
and regular cores, can be treated with the same identities. Promoting the
fixed-background treatment to the coupled Einstein--scalar--vector problem,
with the field junction conditions imposed on the moving shell, will test the
stability results against perturbations of the decoupling sector. Deriving
the stationary extension from an axisymmetric decoupling will turn the
rotating ansatz into a solution of a definite theory. Finally, the linearized
field equations around the slowly rotating configuration, with linearized
junction conditions that include the displacement of the shell and outgoing
radiation at both ends, will replace the free-precession model by a fully
relativistic calculation of the gravitational-wave signal. Together, these
steps would turn the asymmetric wormhole into a concrete benchmark for
confronting exotic compact objects with gravitational-wave observations.

\section*{Acknowledgments}
The work of Kazuharu Bamba was supported in part by the JSPS KAKENHI Grant No.~25KF0176 and a grant-in-aid of academic research of Yamaguchi Scholarship Foundation.

\appendix

\section{Inertia tensor of the rotating shell}
\label{app:inertia}

This appendix collects the expansion of the Newtonian inertia tensor of the
rotating shell used in Sec.~\ref{sec:rotation}. We expand the density
measured by the zero-angular-momentum observers to second order in the spin
and integrate it over the oblate spheroid that represents the throat, first
with the area element of the flat spheroid and then with the proper area
element induced on the junction surface. Writing both integrals with their
angular weights makes explicit which terms fix the ellipticity, how the
deformation factors out of every ratio, and how the result depends on the
representation of the throat. The same decomposition allows a term-by-term
reconstruction of the integrals of Ref.~\cite{Kim2023}, which identifies the
origin of the difference with the ellipticity quoted there.

To second order in $\zeta\equiv j^2/a_0^2$, the density~\eqref{eq:sigmaKS} multiplied
by $\sqrt{\Sigma_0}$ has the form
\begin{equation}
 \sqrt{\Sigma_0}\,\sigma=\sigma_\ast\Big[1+\zeta\left(P+Q\cos^2\vartheta\right)\Big]+O(\zeta^2),
 \qquad
 \sigma_\ast=-\frac{(1+q_0)\sqrt{A_0}}{4\pi},
 \quad
 P=\frac{6-x_0}{2x_0(x_0-2)},
 \quad
 Q=\frac{3-x_0}{2x_0}.
 \label{eq:sigmaexp}
\end{equation}
For the spheroid of Sec.~\ref{sec:rotation}, $r^2\delta_{ij}-x_ix_j$ gives
$y^2+z^2=(a_0^2+j^2)\sin^2\vartheta\sin^2\varphi+a_0^2\cos^2\vartheta$ for $I_1$ and
$x^2+y^2=(a_0^2+j^2)\sin^2\vartheta$ for $I_3$.  The flat area element of this
spheroid is
$\dd A_{\rm f}=\sqrt{\Sigma_0}\sqrt{a_0^2+j^2}\,\sin\vartheta\,\dd\vartheta\,\dd\varphi$,
so that $\sigma\,\dd A_{\rm f}=\sigma_\ast[1+\zeta(P+Q\cos^2\vartheta)]
\sqrt{a_0^2+j^2}\,\sin\vartheta\,\dd\vartheta\,\dd\varphi$.  The angular integrals
needed are
\begin{equation}
 \frac{\int_0^\pi\cos^2\vartheta\sin^3\vartheta\,\dd\vartheta}{\int_0^\pi\sin^3\vartheta\,\dd\vartheta}=\frac15,
 \qquad
 \frac{\int_0^\pi\cos^4\vartheta\sin\vartheta\,\dd\vartheta}{\int_0^\pi\cos^2\vartheta\sin\vartheta\,\dd\vartheta}=\frac35,
 \label{eq:weights}
\end{equation}
the first for the $\sin^2\vartheta$ terms and the second for the $\cos^2\vartheta$
term of $I_1$.  Integrating over $\varphi$ and $\vartheta$,
\begin{equation}
 I_1^{\rm f}=\frac{8\pi}{3}\sigma_\ast a_0^3\left[1+\zeta\left(1+P+\frac{2Q}{5}\right)\right],
 \qquad
 I_3^{\rm f}=\frac{8\pi}{3}\sigma_\ast a_0^3\left[1+\zeta\left(\frac32+P+\frac{Q}{5}\right)\right],
 \label{eq:I13flat}
\end{equation}
where the $\cos^2\vartheta$ term enters $I_1$ with the average $2/5$ of the
weights $1/5$ and $3/5$, and $I_3$ with the weight $1/5$.  The isotropic
coefficient $P$ cancels in the ellipticity, which is
$\epsilon_I^{\rm f}=\zeta(1/2-Q/5)=3(2x_0-1)\zeta/(10x_0)$.

The proper area element of $r=a_0$ differs from the flat one by the term
$2Ma_0j^2\sin^2\vartheta/\Sigma_0$ in $h_{\varphi\varphi}$,
$\dd A=\dd A_{\rm f}\,[1+\zeta\sin^2\vartheta/x_0+O(\zeta^2)]$.  With the weights
\begin{equation}
 \frac{\int_0^\pi\sin^5\vartheta\,\dd\vartheta}{\int_0^\pi\sin^3\vartheta\,\dd\vartheta}=\frac45,
 \qquad
 \frac{\int_0^\pi\cos^2\vartheta\sin^3\vartheta\,\dd\vartheta}{\int_0^\pi\cos^2\vartheta\sin\vartheta\,\dd\vartheta}=\frac25,
\end{equation}
this adds $3\zeta/(5x_0)$ to the bracket of $I_1$ and $4\zeta/(5x_0)$ to that of
$I_3$, so that $\epsilon_I=\epsilon_I^{\rm f}+\zeta/(5x_0)=(6x_0-1)\zeta/(10x_0)$, and
$I_1-I_3=-\epsilon_II_1$ gives Eq.~\eqref{eq:inertia}.  The factor
$1+q_0$ of $\sigma_\ast$ multiplies all moments and drops out of every ratio.
We have checked both results against a direct numerical integration of the
exact density with the exact area elements.

The area element used in Ref.~\cite{Kim2023} is the flat one, written there as
$\sqrt{b^2+a^2\cos^2\theta}\,\sqrt{b^2+a^2}\,\sin\theta$ with $b=a_0$, and its
expansion of the density has the form~\eqref{eq:sigmaexp} with the same $Q$
and with $P_{\rm K}=(x_0+2)/[2x_0(x_0-2)]$, which differs from $P$ by $1/x_0$ and
does not affect the ellipticity.  With this $P_{\rm K}$, the brackets reported
there are
\begin{equation}
 \frac{16x_0^2-10x_0-4}{20x_0(x_0-2)}=1+P_{\rm K}+\frac{2Q}{5},
 \qquad
 \frac{29x_0^2-45x_0+14}{20x_0(x_0-2)}=\frac32+P_{\rm K}+\frac{Q}{10}.
 \label{eq:Kimbrackets}
\end{equation}
The first coincides with Eq.~\eqref{eq:I13flat}; the second corresponds to a
weight $1/10$ instead of $1/5$ for the $\cos^2\vartheta$ term in $I_3$, and leads
to $(13x_0-9)\zeta/(20x_0)$.  With the weight~\eqref{eq:weights} the same
expansion gives $3(2x_0-1)\zeta/(10x_0)$, the flat-element value above.

\bibliographystyle{apsrev4-2}
\bibliography{biblio}

\end{document}